\documentclass[referee]{aa_modi}
\def\targ{IRAS~21078$+$5211}

\def\nh3{NH$_{3}$}
\def\kms{km~s$^{-1}$}
\def\kmsau{km~s$^{-1}$~au$^{-1}$}

\def\Vlsr{$V_{\rm LSR}$}
\def\Jyb{Jy~beam$^{-1}$}

\def\G24{G24.78$+$0.08}

\newcommand{\ms}{$M_{\odot}$}
\newcommand{\ls}{$L_{\odot}$}

\newcommand{\pas}{$\rlap{.}^{\prime\prime}$}
\newcommand{\pss}{$\rlap{.}^{\rm s}$}

\newcommand{\degree}{$^{\circ}$}

\usepackage[numbers,super]{natbib}
\usepackage{lscape}

\usepackage{lineno}
\usepackage{amsmath}

\begin{document}


\title{Snapshot of a magnetohydrodynamic disk wind traced by water maser observations}
    

 \titlerunning{Snapshot of a magnetohydrodynamic disk wind}
   
   \author{L. Moscadelli$^{*}$\inst{1}
          \and
          A. Sanna\inst{2,3}
          \and
          H. Beuther\inst{4}
          \and
          G.A. Oliva\inst{5,6}
          \and
          R. Kuiper\inst{7}
}
  
   \institute{INAF-Osservatorio Astrofisico di Arcetri, Largo E. Fermi 5, 50125 Firenze, Italy \\
             \email{luca.moscadelli@inaf.it}
            \and
            INAF-Osservatorio Astronomico di Cagliari, Via della Scienza 5, 09047, Selargius, CA, Italy 
            \and
            Max-Planck-Institut f\"{u}r Radioastronomie, Auf dem H\"{u}gel 69, D-53121 Bonn, Germany
            \and
            Max Planck Institute for Astronomy, K\"onigstuhl 17, D-69117 Heidelberg, Germany 
            \and 
            Institut für Astronomie und Astrophysik, Universität Tübingen, Auf der Morgenstelle 10, D-72076 Tübingen, Germany
            \and
            Space Research Center (CINESPA), School of Physics, University of Costa Rica, 11501 San Jos\'{e}, Costa Rica
            \and 
            Faculty of Physics, University of Duisburg-Essen, Lotharstraße 1, D-47057 Duisburg, Germany
             }

   \date{}









\abstract{}


\maketitle

{\bf The formation of astrophysical objects of different nature and size, from black holes to gaseous giant planets, involves a disk-jet system, where the disk drives the mass accretion onto a central compact object and the jet is a fast collimated ejection along the disk rotation axis. Magnetohydrodynamic disk winds can provide the link between mass accretion and ejection, which is essential to ensure that the excess angular momentum is removed from the system and accretion onto the central object can proceed. 
However, up to now, we have been lacking direct observational proof of disk winds.
%
This work presents a direct view of the velocity field of a disk wind around a forming massive star.
Achieving a very high spatial resolution of $\approx$~0.05~au, our water maser observations trace the velocities of individual streamlines emerging from the disk orbiting the forming star. We find that, at low elevation above the disk midplane, the flow co-rotates with its launch point in the disk, in agreement with magneto-centrifugal acceleration where the gas is flung away along the magnetic field line anchored to the disk. Beyond the co-rotation point, the flow rises spiraling around the disk rotation axis along a helical magnetic field.
We have performed (resistive-radiative-gravito-) magnetohydrodynamic simulations of the formation of a massive star starting from the gravitational collapse of a rotating cloud core threaded by a magnetic field. A magneto-centrifugally launched jet develops around the 
forming massive star which has properties matching many features of the maser and thermal (continuum and line) observations of our target.
Our results are, presently, the clearest evidence for a magnetohydrodynamic disk wind, and show that water masers and forming massive stars provide a suitable combination of tracer and environment to allow us studying the disk-wind physics.}

\begin{figure}%
\centering
\includegraphics[width=\textwidth]{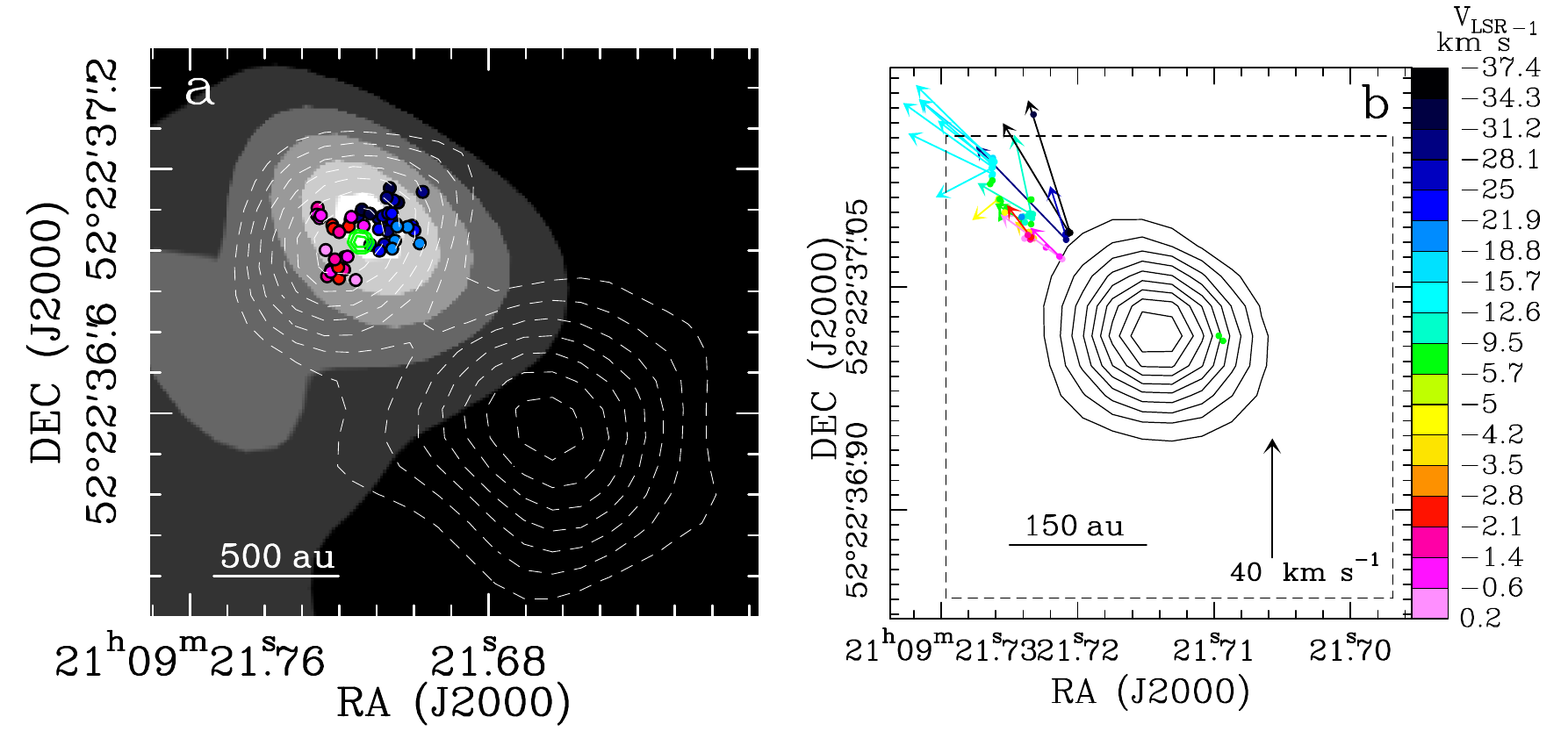}
\caption{Previous NOEMA, JVLA and VLBA observations \cite{Mos16,Mos21} towards \targ. (a)~The gray-scale (from 10 to 35~m\Jyb) map reproduces the NOEMA 1.37~mm continuum emission.
The colored dots represent the channel emission peaks of the CH$_3$CN~$J_K=$ 12$_K$-11$_K$ ($K$ = 3--6) and \ HC$_3$N~$J=$ 24-23 lines, with colors denoting the channel \Vlsr: blue for \ [$-$13.6, $-$10] and red \ [$-$2.5, 0.5]~\kms. The green (70\%, 80\%, and 90\% of \ 0.50~m\Jyb) and white (from 30\% \ to  90\%, in steps of 10\% of \ 0.096~m\Jyb) contours show the JVLA A-Array continuum at 1.3~cm and  5~cm, respectively.
\ (b)~Colored dots and arrows give absolute positions and proper motions of the 22~GHz water masers determined with multi-epoch (2010--2011) VLBA observations, with colors denoting the maser \Vlsr. The black contours (from 10\% \ to  90\%, in steps of 10\% of \ 0.50~m\Jyb) indicate the JVLA A-Array continuum at 1.3~cm. The dashed rectangle delimits the field of view plotted in Fig.~\ref{glo}a.}
\label{NV}
\end{figure}

Magnetohydrodynamic (MHD) disk winds have been proposed to be the engines of the powerful jets observed at varying length scales in many diverse sources, from young stellar objects \cite{Pud83} (YSO) to black holes \cite{Bla82}. 
According to the classical model of an ideal MHD disk wind \cite{Bla82}, in the reference frame co-rotating with the launch point,
the flow streams along the magnetic field line anchored to the accretion disk. An observer at rest sees magneto-centrifugal acceleration: the magnetic field keeps the flow in co-rotation with its launch point while its radial distance increases, till reaching the Alfv\'en point where the poloidal kinetic and magnetic energies are equal. Beyond the Alfv\'en point, the flow spirals outward along the rotation axis with a stably increasing ratio of the streaming onto the rotational velocity, until it gets eventually collimated into a fast jet \cite{Ouy97,Kra99}. 
So far, the best observational evidence for a MHD disk wind has been the finding of line of sight velocity gradients transversal to the jet axis, which are interpreted in terms of jet rotation and the imprint of the magneto-centrifugal acceleration  \cite{Bac02,Hir17,Lee17,Aal20}. However, that is an indirect evidence and the derivation of key parameters, as the launch radius and the magnetic lever arm, can be seriously affected by systematic biases  \cite{Tab20}. On scales of \ $\sim$100~au, a few studies based on Very Long Baseline Interferometry (VLBI) maser (the laser equivalent at microwave band) observations have revealed rotating disk-like\cite{Mat10,Mos14,San15}, conical\cite{Mos11a} or cylindrical \cite{Bur15} maser distributions at the jet root, but the streamlines of a disk wind have been never traced, up to now.

\targ \ is a star forming region of high bolometric luminosity, 
\ 5$\, \times \,$10$^{3}$~\ls \ \cite{Mos16} at a distance of \ 1.63$\pm$0.05~kpc \cite{Xu13}, and harbors a YSO significantly more massive than the Sun. On scales of a few 100~au, by employing the NOrthern Extended Millimeter Array (NOEMA), a disk \cite{Mos21} is observed in high-density molecular tracers (CH$_3$CN and HC$_3$N, see Fig.~\ref{NV}a) rotating around a YSO of mass of \ $5.6\pm2$~\ms. Interferometric observations at radio wavelengths (5~cm) using the Jansky Very Large Array (JVLA) have revealed a jet \cite{Mos16} directed NE-SW (PA $\approx$ 44\degree) emerging from the YSO, whose position at the center of the disk is pinpointed by compact thermal emission observed with the JVLA at 1.3~cm. During 2010--2011 we have performed multi-epoch Very Long Baseline Array (VLBA) observations of the maser emission of the water molecule at 22~GHz. These observations have discovered a cluster of masers placed $\approx$100~au NE from the YSO, whose proper motions are collimated NE-SW (PA = 49\degree) and trace the base of the jet from the YSO \cite{Mos16} (see Fig.~\ref{NV}b). The analysis of the three-dimensional (3D) maser motions, specifically the local standard of rest (LSR) velocity (\Vlsr) gradient transversal to the jet axis and the constant ratio between the toroidal and poloidal velocities, suggested that the jet could be launched from a MHD disk wind.

\begin{figure}%
\centering
\includegraphics[width=\textwidth]{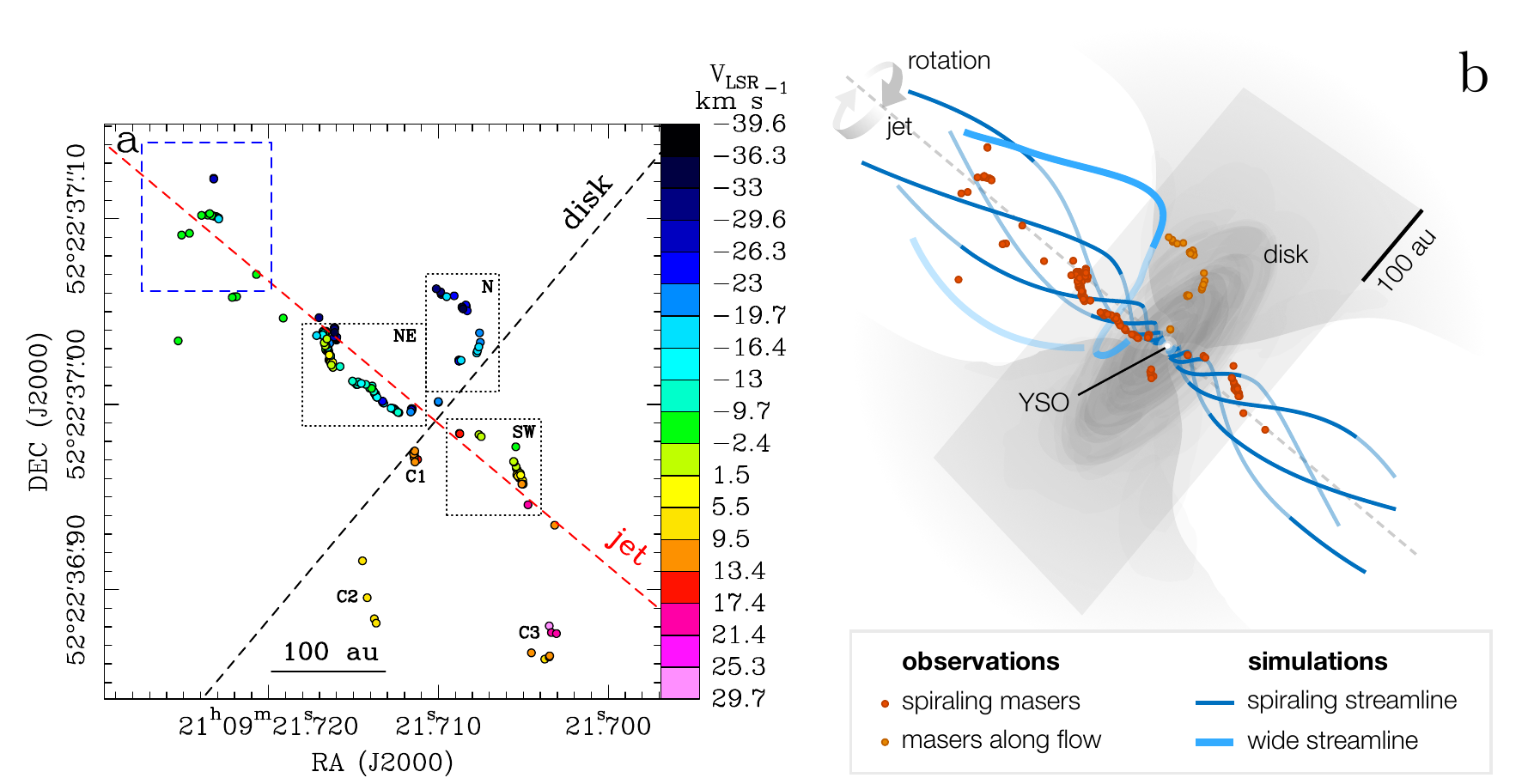}
\caption{October 2020, global VLBI observations of the 22~GHz water masers and 3D view of the proposed interpretation for the maser kinematics. \ (a)~Colored dots give absolute positions of the 22~GHz water masers, with colors denoting the maser \Vlsr. The black dotted rectangles encompass the three regions, to the N, NE and SW, where maser emission concentrate. The blue dashed rectangle delimits the area of maser emission in the previous VLBA observations, and three distinct maser clusters S of the YSO are labeled ``C1'', ``C2'', and ``C3'' (see Appendix~\ref{met_NES}). The red and black dashed lines mark the sky-projected jet and disk axis, respectively.\ (b)~Observed maser positions (red and orange dots) overlaid on top of streamlines (blue lines) computed from resistive-radiative-gravito-MHD simulations of a jet around a forming massive star (see Appendix~\ref{met_simu}). The streamlines close to the rotation axis show significant spiraling motion, in agreement with the kinematic signature of the masers observed in the NE and SW regions. The wide streamline from the simulation illustrates the outflowing trajectory of material from the outer disk, similar to the observed masers in the N region. For context, the protostar, the disk and the outflow cavity have been sketched in gray, based on the density structure obtained in the simulations (see Fig.~\ref{fig: simulation-results}).
 }
\label{glo}
\end{figure}

In October 2020, we performed novel observations (see Fig.~\ref{glo}a) of the water maser emission in \targ\ by including all telescopes available in the VLBI network, with the aim to simulate next-generation radio interferometers which will improve current sensitivities by more than an order of magnitude 
(see Appendix~\ref{met_obs}). In the following, we will show that these new observations prove that the water masers trace magnetized streams of gas emerging from the YSO's disk (see Fig.~\ref{glo}b and Appendix~\ref{met_simu}). The maser emission concentrates in three regions to NE, N, and SW, inside the three dotted rectangles of Fig.~\ref{glo}a. Along the jet axis (the dashed red line 
in Fig.~\ref{glo}a), whose sky-projection is known from previous observations of the maser proper motions and radio jet (see Appendix~\ref{met_shock}), we observe two elongated structures, blue-~and~red-shifted (with respect to the systemic \Vlsr \ of the YSO: $V_{\rm sys} = -6.4$~\kms) to NE and SW, respectively. These structures are the opposite lobes of a collimated outflow from the YSO, located in between the two lobes; the disk axis (the black dashed line in Fig.~\ref{glo}a) is the intercept of the jet axis at the YSO position. From previous VLBA observations we know that the jet axis has to lay close to the plane of the sky, with an inclination \ $\le$~30\degree. According to the maser \Vlsr, the jet is inclined towards us to NE, and away from us to SW.

\begin{figure}%
\centering
\includegraphics[width=0.5\textwidth]{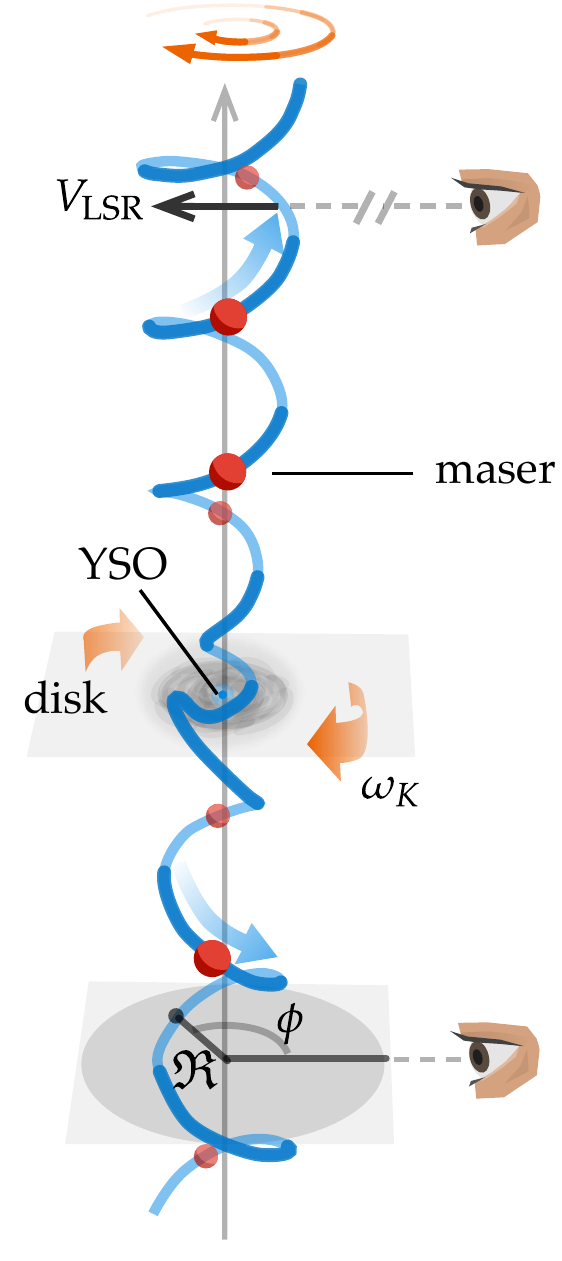}
\caption{Illustration of the spiral motion of the masers (small red balls) along a magnetic field line (blue) emerging from the YSO's disk. The rotation angle, $\phi$, and radius, $\mathfrak{R}$, are labeled. The field line is anchored to a point of the disk rotating at an angular velocity \ $\omega_{\rm K}$. This plot is based on the data from (resistive-radiative-gravito-) MHD simulations of a jet around a forming massive star (see Appendix~\ref{met_simu}).}
\label{ske_1}
\end{figure}

The jet and disk axes provide a convenient coordinate system to refer the maser positions to. In the following, we present the interpretation of the maser kinematics, which is based on the analysis of the three independent observables:  \ $z$, the elevation above the disk plane (or offset along the jet), $R$, the radial distance from the jet axis (or transversal offset), and the maser \Vlsr. As discussed 
in Appendix~\ref{met_obs}, the accuracy of the maser positions is \ $\approx$~0.05~au, and that of the maser \Vlsr \ $\approx$~0.5~\kms. Without loss of generality, we can express the maser velocities as the sum of two terms, one associated with the toroidal component or rotation around the jet axis, $V_{\rm rot}$, and the other associated with the poloidal component including all the contributions owing to non-rotation, $V_{\rm off}$. Since the jet axis is close to the plane of the sky and we observe the rotation close to edge-on (see Fig~\ref{ske_1}), we can write:
\begin{linenomath*}
\begin{align}
V_{\rm LSR} &= V_{\rm off} + V_{\rm rot} = V_{\rm off} + \omega \ \mathfrak{R} \ \sin(\phi) \label{eq_rot} \\
R &= \mathfrak{R} \ \sin(\phi) \label{eq_R} \\
\phi &= \omega \ t \label{eq_phi} 
\end{align}
\end{linenomath*}
where \ $\phi$ \ is the angle between the rotation radius \ $\mathfrak{R}$ \ and the line of sight, and \ $\omega$ \ and \ $t$ \ are the angular velocity and the time, respectively.

\begin{figure*}%
\includegraphics[width=\textwidth]{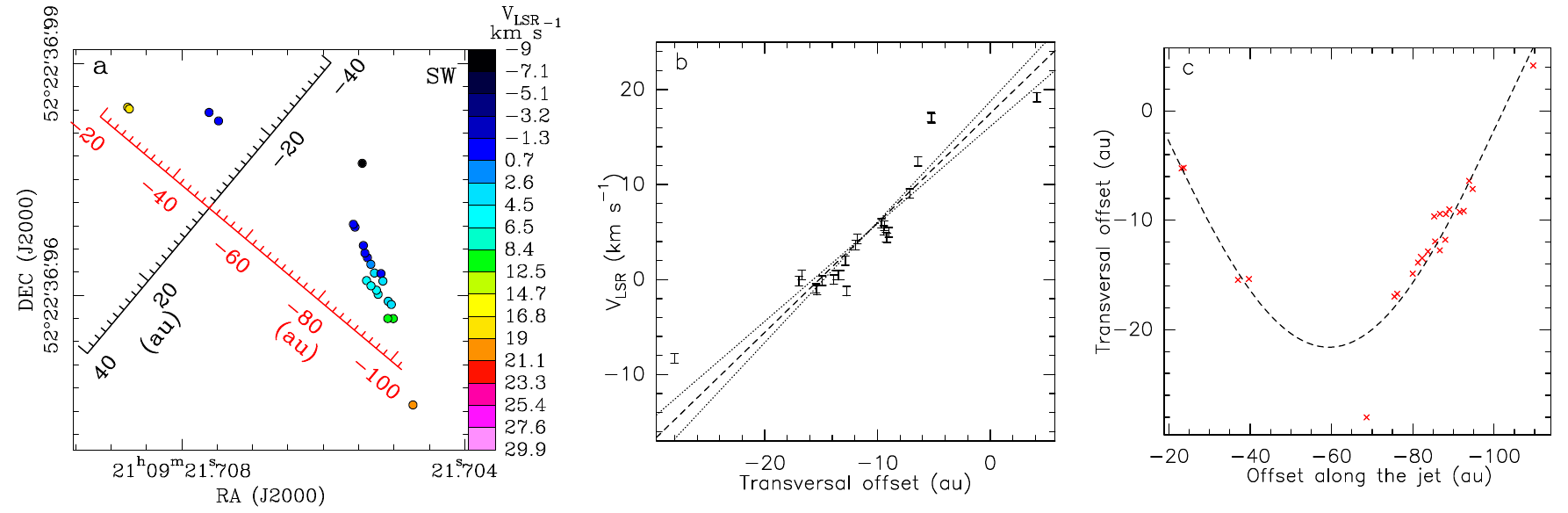}
\caption{The SW spiral motion. (a)~Expanding view of the maser positions and \Vlsr \ in the SW region. Colored dots have the same meaning as in Fig.~\ref{glo}a. 
The distances along the jet (red) and disk (black) axes are indicated. (b)~Plot of the maser \Vlsr \ (and corresponding errors, denoted with errorbars) versus \ $R$. The black dashed and dotted lines show the best linear fit and the associated uncertainty, respectively. (c)~Plot of the maser coordinates \ $R$ \ versus \ $z$. The positional error is smaller than the cross size. The black dashed curve is the fitted sinusoid, whose parameters are reported in Table~\ref{tab_fit}.}
\label{SW_spi}
\end{figure*}

\begin{figure*}%
\centering
\includegraphics[width=\textwidth]{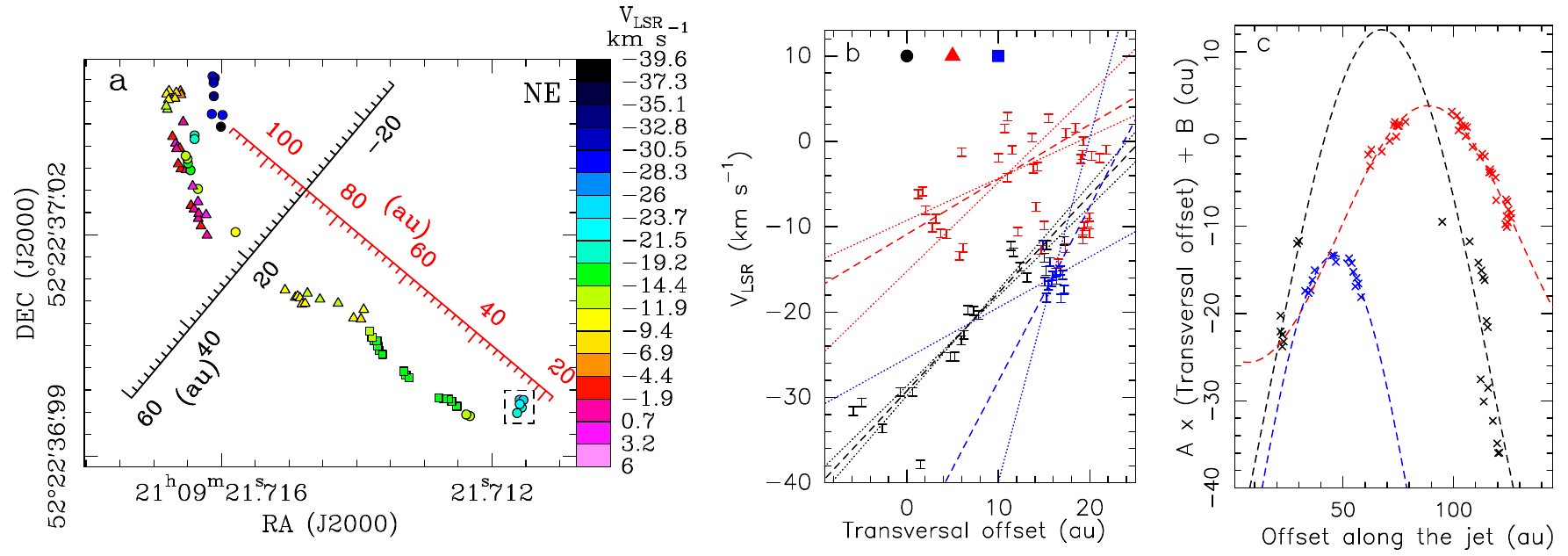}
\caption{The NE spiral motions. (a)~Expanding view of the maser positions and \Vlsr \ in the NE region. Different symbols are employed to identify the three spiral motions: dots for the NE-1, triangles for the NE-2, and squares for the NE-3 spiral motion. Colors denote the maser \Vlsr. The distances along the jet (red) and disk (black) axes are indicated. The black dashed rectangle encompasses the maser cluster closest to the YSO. (b)~In this and following panel, black, red, and blue colors refer to masers belonging to the NE-1, NE-2, and NE-3 streams, respectively. Errorbars, dashed and dotted lines have the same meaning as in Fig.~\ref{SW_spi}b. Masers in the NE-2 stream with similar radii, 10~au $\le R \le 22$~au, but different elevations, 60~au $\le z <$~90~au \ and \ $z \ge 90$~au, have \Vlsr \ different by $\approx$~10~\kms \ (see Appendix~\ref{met_res}). For the NE-2 stream, the linear fit of the \Vlsr\ has been performed considering only the masers with  \ $z \ge 90$~au. (c)~Plot of the linear transformation
of \ $R$ \ versus \ $z$. For each of the three streams, the coefficients $A$ and $B$ are taken equal to the corresponding values of  \ $\omega$ \ and \  $V_{\rm off}$, respectively, in Table~\ref{tab_fit}. We plot the linear transformation of the radii to reduce the overlap and improve the visibility of each of the three streams. The dashed curves are the fitted sinusoids. In Table~\ref{tab_fit}, we report the parameters of the sinusoidal fits of the radii.
}
\label{NE_spi}
\end{figure*}

Fig.~\ref{SW_spi}c shows the remarkable finding that
the spatial coordinates \ $z$ \ and \ $R$ \ of the maser emission in the SW flow satisfy the relation:
\begin{linenomath*}
\begin{align}
R = C \ \sin(f_z (z-z_0)) \label{eq_sin} 
\end{align}
\end{linenomath*}
where \ $C$, the amplitude of the sinusoid, $f_z$, the spatial frequency, and \ $z_0$, the position of zero phase, are fitted constants (see Table~\ref{tab_fit}). 
In Appendix~\ref{met_NE1} we demonstrate that the masers in the NE flow can be separated in three different streams, each of them satisfying the relation~\ref{eq_sin} (see Fig.~\ref{NE_spi}c and Table~\ref{tab_fit}).
The comparison of Eqs.~\ref{eq_R}~and~\ref{eq_sin} leads to a straightforward interpretation of the sinusoidal relation between the coordinates by taking: 1)~$\mathfrak{R} = C $,  and 
\ 2)~$ \phi =  f_z \ |z-z_0|$. The former equation indicates that the rotation radius is the same for all the masers, the latter shows that the motions of rotation around and streaming along the jet axis are locked together, which is the condition for a spiral motion. Denoting with  \ $V_z$ \ the streaming velocity along the jet axis, we can write \ $ |z - z_0| = V_z \ t $ \ and, comparing with  Eq.~\ref{eq_phi}, we derive the relation between the rotation and streaming velocities: 
\begin{linenomath*}
\begin{equation}
 f_z = \omega / V_z \label{eq_fz}
\end{equation}
\end{linenomath*}

According to Eq.~\ref{eq_fz}, the observation of a well defined sinusoidal pattern requires that \  $\omega$ \ and \ $ V_z $ are directly proportional, or constant. The constancy of \ $V_z$ \  implies that \ $ V_{\rm off} $ \ is also constant, because, if the rotation radius does not change, $ V_{\rm off} $ \ is the projection along the line of sight of \ $V_z$.
Following Eqs.~\ref{eq_rot}~and~\ref{eq_R}, the constancy of \  $\omega$ \ and \ $ V_{\rm off} $ \ would result into a tight linear correlation between \Vlsr\ and transversal offsets $R$. While a good linear correlation between \Vlsr\ and \ $R$ \ is observed for the SW and NE-1 spiral motions (see Figs.~\ref{SW_spi}b~and~\ref{NE_spi}b, black symbols), the scatter in velocity is considerable for the NE-2 spiral motion (see Fig.~\ref{NE_spi}b, red symbols). In Appendix~\ref{met_res} we investigate the physical reason for the observation of well defined sinusoidal patterns despite the presence of a significant velocity scatter. Applying the equations of motions for an axisymmetric MHD flow, we find that the magnetic field configuration has to be helical over the maser emission region, and the motion along such an helical field line, in the reference frame co-rotating with the launch point, leads to the sinusoidal pattern of maser positions.

\begin{figure*}%
\centering
\includegraphics[width=\textwidth]{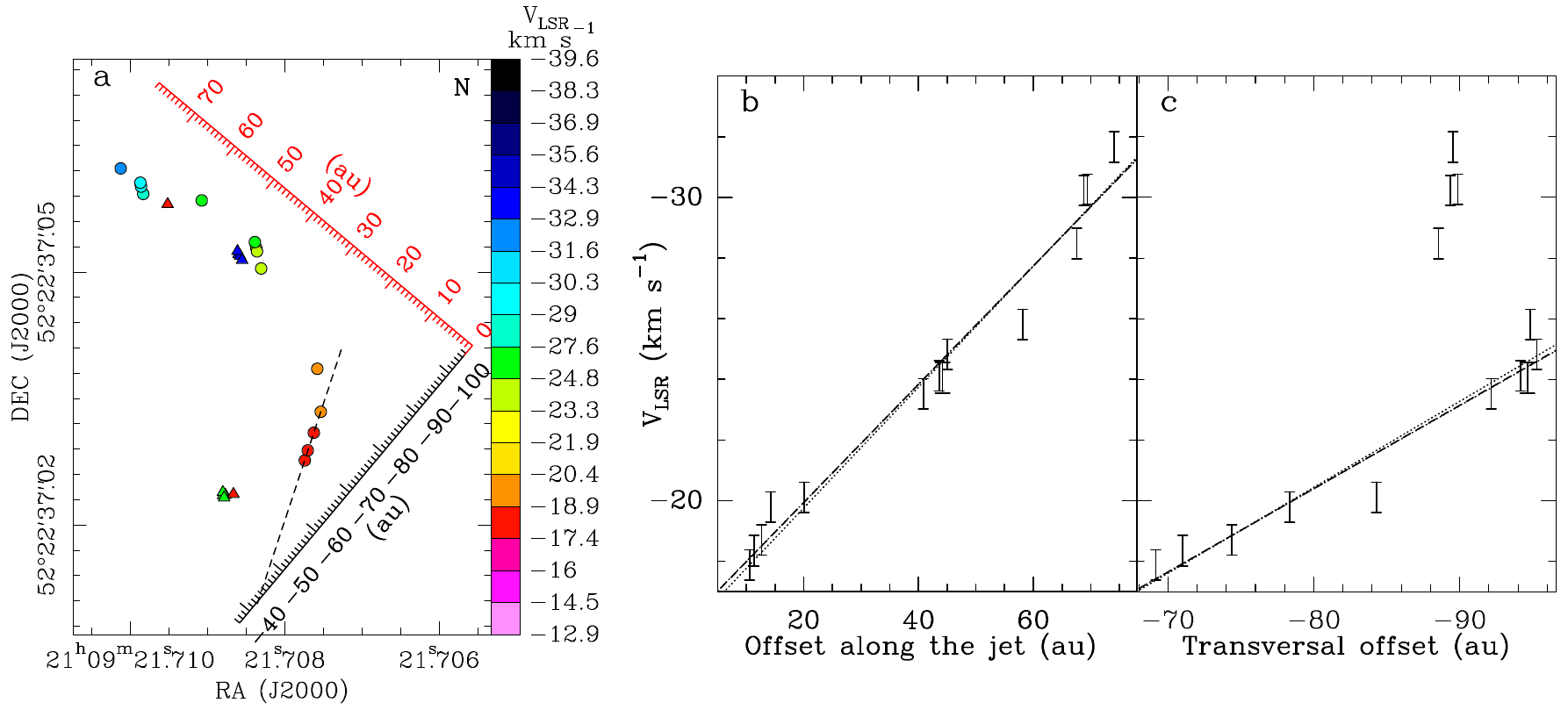}
\caption{The co-rotating N~stream. (a)~Expanding view of the maser positions and \Vlsr \ in the N~region. Colored dots and triangles give absolute positions of the 22~GHz water masers, with colors denoting the maser \Vlsr. The triangles mark a few masers with detached \Vlsr\ excluded from the kinematical analysis. The distances along the jet (red) and disk (black) axes are indicated. The black dashed line shows the linear fit of the positions of the masers at elevation \ $z <$~20~au. \ Plot of maser \Vlsr\ versus $z$~(b) \ and $R$~(c). In both panels, errorbars give the maser \Vlsr\ and corresponding errors, and the black dashed and dotted lines show the best linear fit and the associated uncertainty, respectively. The linear fit of \ \Vlsr\ versus $R$ \ has been performed considering only the masers with \Vlsr\ $\ge$~$-27$~\kms.}
\label{N_wire}
\end{figure*}

We consider now the N region (see Fig.~\ref{glo}a) and show that, in this region as well, the maser kinematics is consistent with the predictions for a MHD disk wind. The N masers have a larger separation from the jet axis than the NE and SW masers. 
A few nearby masers show quite different \Vlsr, which could hint at distinct streams, as observed (see Fig.~\ref{NE_spi}a) and discussed (see Appendix~\ref{met_NE1}) for the NE flow. In this case, however, only a single stream is reasonably well sampled in position and \Vlsr\ with the masers, and we focus our kinematical analysis on that.
The spatial distribution of this stream presents an arc-like shape: a subset of maser features draws a line at small angle with the disk axis and another group extends at higher elevations about parallel to the jet axis (see Fig.~\ref{N_wire}a). Fig.~\ref{N_wire}c shows that the maser \Vlsr \ increases linearly with  $R$  in the range \ $-95$~au $\lesssim  R \lesssim$~$-70$~au \ up to an elevation \ $z \approx$~60~au. The relatively large separation from the jet axis and radial extent, and the arc-like shape of the maser distribution lead us to think that the N emission is observed close to the plane of the sky. In this case, the maser \Vlsr \ should mainly trace rotation, which is also expected to be the dominant velocity component at low elevations above the disk. Then, the good linear correlation between  \Vlsr \ and \ $R$ \ indicates that the masers co-rotate at the same angular velocity, $\omega_{\rm N} = 0.274\pm0.005$~\kmsau, up to \ $R \approx$~$-95$~au \ and \ $z \approx$~60~au. A simple interpretation is in terms of a magneto-centrifugally accelerated stream of gas emerging from a point of the disk. A disk in Keplerian rotation around an YSO of about \ 5.6~\ms \ attains an angular velocity equal to $\omega_{\rm N} $ at \ $R \approx$~40~au. The line drawn by the masers at the lowest elevations intercepts the disk axis close to $-40$~au (see Fig.~\ref{N_wire}a), as expected if the gas, launched from the disk, first streams approximately along a straight line and then progressively bends up along the jet axis. Note that, basing on
 Eqs.~\ref{eq_rot}~and~\ref{eq_R}, the derivation of \ $\omega_{\rm N} $ \ does not depend on the maser geometry. Therefore, the finding that the masers lay  along a line intercepting the disk at \ $\approx$~$-40$~au provides an "a posteriori" test of the assumption that the N emission is observed close to the plane of the sky. 
 
 The masers found at elevation \ $z$ $>$~60~au appear to set aside of the arc-like distribution (see Fig.~\ref{N_wire}a) and their \Vlsr\ are significantly more negative and do not follow the linear correlation with the radius 
 (see Fig.~\ref{N_wire}c). A natural interpretation is that the location at  \ $R \approx$~$-95$~au \ and \ $z \approx$~60~au \ is the Alfv\'en point of the stream, and beyond that the gas is not any longer co-rotating. The lever arm of the stream is \ $\approx$~(95~au)/(40~au) = $2.4$, in agreement with the values of \ 2--3 \ predicted by theory \cite{Pud07,Pud19}. The more negative \Vlsr\ are explained if  \ $V_z$ (and the absolute value of \ $V_{\rm off}$) increases with the elevation, as a consequence of the magneto-centrifugal launching. The linear correlation between  \Vlsr\ and \ $z$ \ (${\rm d} V_{\rm LSR} / {\rm d} z = -0.195\pm0.002$~\kmsau) shown in  Fig.~\ref{N_wire}b results from the combination of the two regimes: 1)~sub-Alfv\'enic, where \ $V_{LSR} \propto R$ \ and \ the gas streams approximately along a straight line, that is \ 
\ $R \propto z$; \ 2)~trans-Alfv\'enic, where \ $V_z$ \ increases fast with \ $z$ \ and \ $V_{\rm off}$ starts to be significant.

From the previous analysis, a MHD disk wind seems to be a natural frame to explain both the spiral motions traced by the masers close to the jet axis in the NE and SW regions and the gas co-rotation along the N stream.
If some locations of the YSO's disk are perturbed, the flow emerging from those perturbed launch points should harbor internal shocks (see Appendix~\ref{met_shock}), which travel outward along spiraling trajectories. These internal shocks provide physical conditions suitable for the excitation of the water masers \cite{Eli92, Hol13, Kau96}. The spiral motions traced with the masers would correspond to portions of the trajectories beyond the Alfv\'en point where the rotation radius keeps about constant. An essential feature of the proposed model is that, as the launch point rotates, the maser emissions have to travel along spatially distinct, spiraling trajectories. These trajectories are invariant under rotation and nearby masers share the same orbital parameters. However, since the masers sample different trajectories, we need to make a distinction between the angular velocity of the trajectory, $\omega$, derived  through the linear fit of the maser \Vlsr \ versus  \ $R$ (see Eqs.~\ref{eq_rot}~and~\ref{eq_R}, and Figs.~\ref{SW_spi}b~and~\ref{NE_spi}b), and the effective angular velocity of rotation, $\omega_{\rm e}$, the one to be used in Eqs.~\ref{eq_phi}~and~\ref{eq_fz}. Since the different trajectories are rigidly anchored to the launch point and water masers at higher (absolute) elevations have been launched earlier in time, the simple relation holds:
\begin{linenomath*}
\begin{equation}
 \omega_{\rm e} = \omega - \omega_{\rm K} \label{eq_we}
\end{equation}
\end{linenomath*}
where \ $\omega_{\rm K}$ \ is the Keplerian angular velocity of the launch point. Eq.~\ref{eq_we} shows that \ $ \omega_{\rm e} $ \ is the angular velocity of the spiraling trajectory as observed in the reference frame co-rotating with the launch point.
Based on axisymmetric MHD models \cite{Pese04,Tab20}, the ratio \ $\omega_{\rm K} / \omega$ \ increases stably from 1 up to a value $\approx$~4 while the gas climbs from \ $z_{\rm A}$ \  to \ $10 \ \mathfrak{R}_{\rm K}$, where  \ $z_{\rm A}$ \ is the elevation of the Alfv\'en point and  \ $\mathfrak{R}_{\rm K}$ \ is the launch radius.
Being \ $\omega \le \omega_{\rm K}$, the negative value of \ $ \omega_{\rm e}$ \ indicates that the rotation angle of the maser positions decreases with \ $z$.
Following the previous discussion, Eq.~\ref{eq_fz} has to be corrected by replacing \ $\omega$ \ with \ $\omega_{\rm e}$:
\begin{linenomath*}
\begin{equation}
 f_z = |\omega_{\rm e}| / V_z \label{eq_fz2}
\end{equation}
\end{linenomath*}

A good test of the above considerations comes directly from our data. Assuming that all the masers move along a single trajectory and using the fitted values of \ $\omega$ \ and \ $f_z$ \ in Eq.~\ref{eq_fz}, we obtain implausibly small values for \ $V_z$ :  \ 31, 37, 17 and 49~\kms, for the SW, NE-1, NE-2, and NE-3 spiral motions, respectively. There are two strong observational evidences that the derived streaming velocities are too small. First, comparing them with the values of \ $V_{\rm off}$ (see Table~\ref{tab_fit}), the two velocities have similar amplitudes, and that is inconsistent with the expectation that  \ $V_{\rm off}$ \ be the line of sight component of \ $V_z$ \ and the jet axis being close to the plane of the sky.  Second, taking the ratio between the highest elevations reached by the masers, that is \ 100--130~au (see Figs.~\ref{SW_spi}~and~\ref{NE_spi}), and the above values of \ $V_z$, the derived traveling times of \ 15--40~yr exceed the separation of 10~yr since the previous VLBA observations, when no maser emission was detected at corresponding positions. 

Since \ $V_{\rm off}$ \ corresponds to the line of sight projection of \ $V_z$, we can write:
\begin{linenomath*}
\begin{align}
V_z = |V_{\rm off} - V_{\rm sys}| \ / \ \sin(i_{\rm axi}) \label{eq_axi}
\end{align}
\end{linenomath*}
where \ $V_{\rm off}$ \ is corrected for the systemic \Vlsr \ of the YSO, and \ $ i_{\rm axi}$ \ is the inclination angle of the jet axis with the plane of the sky. As we know that \ $ i_{\rm axi} \le 30^{\circ}$, Eq.~\ref{eq_axi} allows us to derive a lower limit for \ $V_z$, reported in Table~\ref{tab_der}.  
Using the derived lower limit of \ $V_z$ \ and the corresponding value of \ $f_z$ (see Table~\ref{tab_fit}), by means of Eq.~\ref{eq_fz2} we can calculate a lower limit for \ $\omega_{\rm e}$. Finally, we use Eq.~\ref{eq_we} and the fitted value of \ $\omega$ (see Table~\ref{tab_fit}) to infer a lower limit for \ $  \omega_{\rm K} = \omega + |\omega_{\rm e}|$ \ and, knowing the mass of the YSO, a corresponding upper limit for the launch radius  \ $\mathfrak{R}_{\rm K}$ (see Table~\ref{tab_der}).
The NE-1 stream, which extends the most in elevation (from 20 to 130~au, 
see Fig.~\ref{NE_spi}a), includes a group of maser features at elevation of \ $\approx$~20~au, which should be located closer to the Alfv\'en point. In Appendix~\ref{met_wk}, we study the change of \Vlsr \ versus  \ $R$ \ internal to this cluster and obtain a lower limit for \ $\omega_{\rm K} \ge 2.4$~\kmsau \ that agrees well with the corresponding value reported in Table~\ref{tab_der} for the NE-1 stream. The results shown in Table~\ref{tab_der} are in general agreement with the theoretical predictions from MHD disk winds \cite{Pud07,Pud19}: the inferred lower limits on the streaming velocity \ $V_z$ \ increase steeply with decreasing launch radius;  
for the SW and NE-3 streams, which are those with relatively lower uncertainty in the measurement of both \ $\omega$ \ and \  $\mathfrak{R}$ (see Table~\ref{tab_fit}), the ratio between  \ $V_z$ \ and  the rotation velocity \ $\omega \  \mathfrak{R}$ \ is  \ $\ge$~2--3.


 
 In conclusion, our observations resolve, for the first time, the kinematics of a MHD disk wind on length scales of \ 1--100~au, allowing us to study the velocity pattern of individual streamlines launched from the disk. As represented in Fig.~\ref{glo}b, close to the disk rotation axis we observe flows spiraling outward along a helical magnetic field, launched from locations of the disk at radii \ $\le$~6--17~au. At larger separation from the rotation axis, we observe a stream of gas co-rotating with its launch point from the disk at radius of $\approx$~40~au, in agreement with the predictions for magneto-centrifugal acceleration. 
Our interpretation is supported by (resistive-radiative-gravito-) MHD simulations of the formation of a massive star that lead to a magneto-centrifugally launched jet whose properties agree with our maser and thermal (continuum and line) observations of \targ.
 These results provide the best evidence for a MHD disk wind to date. Since water maser emission is widespread in YSOs, sensitive VLBI observations of water masers can be a valuable tool to investigate the physics of disk winds.

\section*{Acknowledgments}
We thank Christian Fendt and Daniele Galli for very useful discussion.
GAO acknowledges financial support from the Deutscher Akademischer Austauschdienst (DAAD), under the program Research Grants - Doctoral Projects in Germany, and complementary financial support for the completion of the Doctoral degree by the University of Costa Rica, as part of their scholarship program for postgraduate studies in foreign institutions.
HB acknowledges support from the European Research Council under the Horizon 2020 Framework Programme via the ERC Consolidator Grant CSF-648505. HB also acknowledges support from the Deutsche Forschungsgemeinschaft in the Collaborative Research Center (SFB 881) “The Milky Way System” (subproject B1).
RK acknowledges financial support via the Emmy Noether and Heisenberg Research Grants funded by the German Research Foundation (DFG) under grant no.~KU 2849/3 and 2849/9.
 The European VLBI Network is a joint facility of independent European,
    African, Asian, and North American radio astronomy institutes.
    Scientific results from data presented in this publication are 
    derived from the following EVN project code:  GM077.
    
%
   \bibliographystyle{abbrvnat_modi} 
   \bibliography{biblio} 

\begin{thebibliography}{41}
\providecommand{\natexlab}[1]{#1}
\providecommand{\url}[1]{\texttt{#1}}
\expandafter\ifx\csname urlstyle\endcsname\relax
  \providecommand{\doi}[1]{doi: #1}\else
  \providecommand{\doi}{doi: \begingroup \urlstyle{rm}\Url}\fi

\bibitem[{Aalto} et~al.(2020){Aalto}, {Falstad}, {Muller}, {Wada}, {Gallagher},
  {K{\"o}nig}, {Sakamoto}, {Vlemmings}, {Ceccobello}, {Dasyra}, {Combes},
  {Garc{\'\i}a-Burillo}, {Oya}, {Mart{\'\i}n}, {van der Werf}, {Evans}, and
  {Kotilainen}]{Aal20}
S.~{Aalto}, N.~{Falstad}, S.~{Muller}, and others.
\newblock {ALMA resolves the remarkable molecular jet and rotating wind in the
  extremely radio-quiet galaxy NGC 1377}.
\newblock \emph{\aap}, 640:\penalty0 A104, Aug. 2020.
\newblock \doi{10.1051/0004-6361/202038282}.

\bibitem[{Anglada} et~al.(2018){Anglada}, {Rodr{\'\i}guez}, and
  {Carrasco-Gonz{\'a}lez}]{Ang18}
G.~{Anglada}, L.~F. {Rodr{\'\i}guez}, and C.~{Carrasco-Gonz{\'a}lez}.
\newblock {Radio jets from young stellar objects}.
\newblock \emph{\aapr}, 26\penalty0 (1):\penalty0 3, Jun 2018.
\newblock \doi{10.1007/s00159-018-0107-z}.

\bibitem[{Bacciotti} et~al.(2002){Bacciotti}, {Ray}, {Mundt}, {Eisl{\"o}ffel},
  and {Solf}]{Bac02}
F.~{Bacciotti}, T.~P. {Ray}, R.~{Mundt}, and others.
\newblock {Hubble Space Telescope/STIS Spectroscopy of the Optical Outflow from
  DG Tauri: Indications for Rotation in the Initial Jet Channel}.
\newblock \emph{\apj}, 576\penalty0 (1):\penalty0 222--231, Sep 2002.
\newblock \doi{10.1086/341725}.

\bibitem[{Blandford} and {Payne}(1982)]{Bla82}
R.~D. {Blandford} and D.~G. {Payne}.
\newblock {Hydromagnetic flows from accretion disks and the production of radio
  jets.}
\newblock \emph{\mnras}, 199:\penalty0 883--903, June 1982.
\newblock \doi{10.1093/mnras/199.4.883}.

\bibitem[{Burns} et~al.(2015){Burns}, {Imai}, {Handa}, {Omodaka}, {Nakagawa},
  {Nagayama}, and {Ueno}]{Bur15}
R.~A. {Burns}, H.~{Imai}, T.~{Handa}, and others.
\newblock {A `water spout' maser jet in S235AB-MIR}.
\newblock \emph{\mnras}, 453\penalty0 (3):\penalty0 3163--3173, Nov. 2015.
\newblock \doi{10.1093/mnras/stv1836}.

\bibitem[{Caratti o Garatti} et~al.(2017){Caratti o Garatti}, {Stecklum},
  {Garcia Lopez}, {Eisl{\"o}ffel}, {Ray}, {Sanna}, {Cesaroni}, {Walmsley},
  {Oudmaijer}, {de Wit}, {Moscadelli}, {Greiner}, {Krabbe}, {Fischer}, {Klein},
  and {Iba{\~n}ez}]{Car17}
A.~{Caratti o Garatti}, B.~{Stecklum}, R.~{Garcia Lopez}, and others.
\newblock {Disk-mediated accretion burst in a high-mass young stellar object}.
\newblock \emph{Nature Physics}, 13:\penalty0 276--279, Mar. 2017.
\newblock \doi{10.1038/nphys3942}.

\bibitem[{Elitzur} et~al.(1989){Elitzur}, {Hollenbach}, and {McKee}]{Eli89}
M.~{Elitzur}, D.~J. {Hollenbach}, and C.~F. {McKee}.
\newblock {H2O masers in star-forming regions}.
\newblock \emph{\apj}, 346:\penalty0 983--990, Nov. 1989.

\bibitem[{Elitzur} et~al.(1992){Elitzur}, {Hollenbach}, and {McKee}]{Eli92}
M.~{Elitzur}, D.~J. {Hollenbach}, and C.~F. {McKee}.
\newblock {Planar H2O masers in star-forming regions}.
\newblock \emph{\apj}, 394:\penalty0 221--227, July 1992.
\newblock \doi{10.1086/171574}.

\bibitem[{Hirota} et~al.(2017){Hirota}, {Machida}, {Matsushita}, {Motogi},
  {Matsumoto}, {Kim}, {Burns}, and {Honma}]{Hir17}
T.~{Hirota}, M.~N. {Machida}, Y.~{Matsushita}, and others.
\newblock {Disk-driven rotating bipolar outflow in Orion Source I}.
\newblock \emph{Nature Astronomy}, 1:\penalty0 0146, July 2017.
\newblock \doi{10.1038/s41550-017-0146}.

\bibitem[{Hollenbach} et~al.(2013){Hollenbach}, {Elitzur}, and {McKee}]{Hol13}
D.~{Hollenbach}, M.~{Elitzur}, and C.~F. {McKee}.
\newblock {Interstellar H$_{2}$O Masers from J Shocks}.
\newblock \emph{\apj}, 773:\penalty0 70, Aug. 2013.
\newblock \doi{10.1088/0004-637X/773/1/70}.

\bibitem[{Hunter} et~al.(2017){Hunter}, {Brogan}, {MacLeod}, {Cyganowski},
  {Chandler}, {Chibueze}, {Friesen}, {Indebetouw}, {Thesner}, and
  {Young}]{Hun17}
T.~R. {Hunter}, C.~L. {Brogan}, G.~{MacLeod}, and others.
\newblock {An Extraordinary Outburst in the Massive Protostellar System
  NGC6334I-MM1: Quadrupling of the Millimeter Continuum}.
\newblock \emph{\apjl}, 837\penalty0 (2):\penalty0 L29, Mar. 2017.
\newblock \doi{10.3847/2041-8213/aa5d0e}.

\bibitem[{Kaufman} and {Neufeld}(1996)]{Kau96}
M.~J. {Kaufman} and D.~A. {Neufeld}.
\newblock {Water Maser Emission from Magnetohydrodynamic Shock Waves}.
\newblock \emph{\apj}, 456:\penalty0 250--+, Jan. 1996.

\bibitem[{K{\"o}lligan} and {Kuiper}(2018)]{Koe18}
A.~{K{\"o}lligan} and R.~{Kuiper}.
\newblock {Jets and outflows of massive protostars. From cloud collapse to jet
  launching and cloud dispersal}.
\newblock \emph{\aap}, 620:\penalty0 A182, Dec 2018.
\newblock \doi{10.1051/0004-6361/201833686}.

\bibitem[{Krasnopolsky} et~al.(1999){Krasnopolsky}, {Li}, and
  {Blandford}]{Kra99}
R.~{Krasnopolsky}, Z.-Y. {Li}, and R.~{Blandford}.
\newblock {Magnetocentrifugal Launching of Jets from Accretion Disks. I. Cold
  Axisymmetric Flows}.
\newblock \emph{\apj}, 526\penalty0 (2):\penalty0 631--642, Dec. 1999.
\newblock \doi{10.1086/308023}.

\bibitem[{Kuiper} et~al.(2010){Kuiper}, {Klahr}, {Beuther}, and
  {Henning}]{Kui10}
R.~{Kuiper}, H.~{Klahr}, H.~{Beuther}, and T.~{Henning}.
\newblock {Circumventing the Radiation Pressure Barrier in the Formation of
  Massive Stars via Disk Accretion}.
\newblock \emph{\apj}, 722:\penalty0 1556--1576, Oct. 2010.
\newblock \doi{10.1088/0004-637X/722/2/1556}.

\bibitem[{Kuiper} et~al.(2020){Kuiper}, {Yorke}, and {Mignone}]{Kui20}
R.~{Kuiper}, H.~W. {Yorke}, and A.~{Mignone}.
\newblock {Makemake + Sedna: A Continuum Radiation Transport and
  Photoionization Framework for Astrophysical Newtonian Fluid Dynamics}.
\newblock \emph{\apjs}, 250\penalty0 (1):\penalty0 13, Sept. 2020.
\newblock \doi{10.3847/1538-4365/ab9a36}.

\bibitem[{Lee} et~al.(2017){Lee}, {Ho}, {Li}, {Hirano}, {Zhang}, and
  {Shang}]{Lee17}
C.-F. {Lee}, P.~T.~P. {Ho}, Z.-Y. {Li}, and others.
\newblock {A rotating protostellar jet launched from the innermost disk of HH
  212}.
\newblock \emph{Nature Astronomy}, 1:\penalty0 0152, July 2017.
\newblock \doi{10.1038/s41550-017-0152}.

\bibitem[{Machida} et~al.(2007){Machida}, {Inutsuka}, and {Matsumoto}]{Mac07}
M.~N. {Machida}, S.-i. {Inutsuka}, and T.~{Matsumoto}.
\newblock {Magnetic Fields and Rotations of Protostars}.
\newblock \emph{\apj}, 670\penalty0 (2):\penalty0 1198--1213, Dec. 2007.
\newblock \doi{10.1086/521779}.

\bibitem[{Matthews} et~al.(2010){Matthews}, {Greenhill}, {Goddi}, {Chandler},
  {Humphreys}, and {Kunz}]{Mat10}
L.~D. {Matthews}, L.~J. {Greenhill}, C.~{Goddi}, and others.
\newblock {A Feature Movie of SiO Emission 20-100 AU from the Massive Young
  Stellar Object Orion Source I}.
\newblock \emph{\apj}, 708:\penalty0 80--92, Jan. 2010.
\newblock \doi{10.1088/0004-637X/708/1/80}.

\bibitem[{Mignone} et~al.(2007){Mignone}, {Bodo}, {Massaglia}, {Matsakos},
  {Tesileanu}, {Zanni}, and {Ferrari}]{Mig07}
A.~{Mignone}, G.~{Bodo}, S.~{Massaglia}, and others.
\newblock {PLUTO: A Numerical Code for Computational Astrophysics}.
\newblock \emph{\apjs}, 170\penalty0 (1):\penalty0 228--242, May 2007.
\newblock \doi{10.1086/513316}.

\bibitem[{Moscadelli} and {Goddi}(2014)]{Mos14}
L.~{Moscadelli} and C.~{Goddi}.
\newblock {A multiple system of high-mass YSOs surrounded by disks in NGC 7538
  IRS1 . Gas dynamics on scales of 10-700 AU from CH$_{3}$OH maser and NH$_{3}$
  thermal lines}.
\newblock \emph{\aap}, 566:\penalty0 A150, June 2014.
\newblock \doi{10.1051/0004-6361/201423420}.

\bibitem[{Moscadelli} et~al.(2011{\natexlab{a}}){Moscadelli}, {Cesaroni},
  {Rioja}, {Dodson}, and {Reid}]{Mos11a}
L.~{Moscadelli}, R.~{Cesaroni}, M.~J. {Rioja}, and others.
\newblock {Methanol and water masers in IRAS 20126+4104: the distance, the
  disk, and the jet}.
\newblock \emph{\aap}, 526:\penalty0 A66+, Feb. 2011{\natexlab{a}}.
\newblock \doi{10.1051/0004-6361/201015641}.

\bibitem[{Moscadelli} et~al.(2011{\natexlab{b}}){Moscadelli}, {Sanna}, and
  {Goddi}]{Mos11b}
L.~{Moscadelli}, A.~{Sanna}, and C.~{Goddi}.
\newblock {Unveiling the gas kinematics at 10 AU scales in high-mass
  star-forming regions. Milliarcsecond structure of 6.7 GHz methanol masers}.
\newblock \emph{\aap}, 536:\penalty0 A38, Dec. 2011{\natexlab{b}}.
\newblock \doi{10.1051/0004-6361/201117791}.

\bibitem[{Moscadelli} et~al.(2016){Moscadelli}, {S{\'a}nchez-Monge}, {Goddi},
  {Li}, {Sanna}, {Cesaroni}, {Pestalozzi}, {Molinari}, and {Reid}]{Mos16}
L.~{Moscadelli}, {\'A}.~{S{\'a}nchez-Monge}, C.~{Goddi}, and others.
\newblock {Outflow structure within 1000 au of high-mass YSOs. I. First results
  from a combined study of maser and radio continuum emission}.
\newblock \emph{\aap}, 585:\penalty0 A71, Jan. 2016.
\newblock \doi{10.1051/0004-6361/201526238}.

\bibitem[{Moscadelli} et~al.(2020){Moscadelli}, {Sanna}, {Goddi}, {Krishnan},
  {Massi}, and {Bacciotti}]{Mos20}
L.~{Moscadelli}, A.~{Sanna}, C.~{Goddi}, and others.
\newblock {Protostellar Outflows at the EarliesT Stages (POETS). IV.
  Statistical properties of the 22 GHz H$_{2}$O masers}.
\newblock \emph{\aap}, 635:\penalty0 A118, Mar. 2020.
\newblock \doi{10.1051/0004-6361/202037472}.

\bibitem[{Moscadelli} et~al.(2021){Moscadelli}, {Beuther}, {Ahmadi}, {Gieser},
  {Massi}, {Cesaroni}, {S{\'a}nchez-Monge}, {Bacciotti}, {Beltr{\'a}n},
  {Csengeri}, {Galv{\'a}n-Madrid}, {Henning}, {Klaassen}, {Kuiper}, {Leurini},
  {Longmore}, {Maud}, {M{\"o}ller}, {Palau}, {Peters}, {Pudritz}, {Sanna},
  {Semenov}, {Urquhart}, {Winters}, and {Zinnecker}]{Mos21}
L.~{Moscadelli}, H.~{Beuther}, A.~{Ahmadi}, and others.
\newblock {Multi-scale view of star formation in IRAS 21078+5211: from clump
  fragmentation to disk wind}.
\newblock \emph{\aap}, 647:\penalty0 A114, Mar. 2021.
\newblock \doi{10.1051/0004-6361/202039837}.

\bibitem[{Mouschovias} and {Spitzer}(1976)]{Mou76}
T.~C. {Mouschovias} and J.~{Spitzer}, L.
\newblock {Note on the collapse of magnetic interstellar clouds.}
\newblock \emph{\apj}, 210:\penalty0 326, Dec. 1976.
\newblock \doi{10.1086/154835}.

\bibitem[{Oliva} and {Kuiper}(2020)]{Oli20}
G.~A. {Oliva} and R.~{Kuiper}.
\newblock {Modeling disk fragmentation and multiplicity in massive star
  formation}.
\newblock \emph{\aap}, 644:\penalty0 A41, Dec. 2020.
\newblock \doi{10.1051/0004-6361/202038103}.

\bibitem[{Ouyed} and {Pudritz}(1997)]{Ouy97}
R.~{Ouyed} and R.~E. {Pudritz}.
\newblock {Numerical Simulations of Astrophysical Jets from Keplerian Disks. I.
  Stationary Models}.
\newblock \emph{\apj}, 482\penalty0 (2):\penalty0 712--732, June 1997.
\newblock \doi{10.1086/304170}.

\bibitem[{Pelletier} and {Pudritz}(1992)]{Pel92}
G.~{Pelletier} and R.~E. {Pudritz}.
\newblock {Hydromagnetic Disk Winds in Young Stellar Objects and Active
  Galactic Nuclei}.
\newblock \emph{\apj}, 394:\penalty0 117, July 1992.
\newblock \doi{10.1086/171565}.

\bibitem[{Pesenti} et~al.(2004){Pesenti}, {Dougados}, {Cabrit}, {Ferreira},
  {Casse}, {Garcia}, and {O'Brien}]{Pese04}
N.~{Pesenti}, C.~{Dougados}, S.~{Cabrit}, and others.
\newblock {Predicted rotation signatures in MHD disc winds and comparison to DG
  Tau observations.}
\newblock \emph{\aap}, 416:\penalty0 L9--L12, Mar. 2004.
\newblock \doi{10.1051/0004-6361:20040033}.

\bibitem[{Pudritz} and {Norman}(1983)]{Pud83}
R.~E. {Pudritz} and C.~A. {Norman}.
\newblock {Centrifugally driven winds from contracting molecular disks}.
\newblock \emph{\apj}, 274:\penalty0 677--697, Nov. 1983.
\newblock \doi{10.1086/161481}.

\bibitem[{Pudritz} and {Ray}(2019)]{Pud19}
R.~E. {Pudritz} and T.~P. {Ray}.
\newblock {The Role of Magnetic Fields in Protostellar Outflows and Star
  Formation}.
\newblock \emph{Frontiers in Astronomy and Space Sciences}, 6:\penalty0 54,
  July 2019.
\newblock \doi{10.3389/fspas.2019.00054}.

\bibitem[{Pudritz} et~al.(2007){Pudritz}, {Ouyed}, {Fendt}, and
  {Brandenburg}]{Pud07}
R.~E. {Pudritz}, R.~{Ouyed}, C.~{Fendt}, and A.~{Brandenburg}.
\newblock {Disk Winds, Jets, and Outflows: Theoretical and Computational
  Foundations}.
\newblock In B.~{Reipurth}, D.~{Jewitt}, and K.~{Keil}, editors,
  \emph{Protostars and Planets V}, page 277, Jan 2007.

\bibitem[Reid et~al.(1988)Reid, Schneps, Moran, Gwinn, Genzel, Downes, and
  R{\"o}nn{\"a}ng]{Rei88}
M.~J. Reid, M.~H. Schneps, J.~M. Moran, and others.
\newblock The distance to the center of the galaxy - h2o maser proper motions
  in sagittarius b2(n).
\newblock \emph{\apj}, 330:\penalty0 809, 1988.

\bibitem[{Sanna} et~al.(2010){Sanna}, {Moscadelli}, {Cesaroni}, {Tarchi},
  {Furuya}, and {Goddi}]{San10a}
A.~{Sanna}, L.~{Moscadelli}, R.~{Cesaroni}, and others.
\newblock {VLBI study of maser kinematics in high-mass star-forming regions. I.
  G16.59-0.05}.
\newblock \emph{\aap}, 517:\penalty0 A71+, July 2010.
\newblock \doi{10.1051/0004-6361/201014233}.

\bibitem[{Sanna} et~al.(2015){Sanna}, {Surcis}, {Moscadelli}, {Cesaroni},
  {Goddi}, {Vlemmings}, and {Caratti o Garatti}]{San15}
A.~{Sanna}, G.~{Surcis}, L.~{Moscadelli}, and others.
\newblock {Velocity and magnetic fields within 1000 AU of a massive YSO}.
\newblock \emph{\aap}, 583:\penalty0 L3, Nov. 2015.
\newblock \doi{10.1051/0004-6361/201526806}.

\bibitem[{Staff} et~al.(2014){Staff}, {Koning}, {Ouyed}, and {Pudritz}]{Sta14}
J.~{Staff}, N.~{Koning}, R.~{Ouyed}, and R.~{Pudritz}.
\newblock {Three-dimensional simulations of MHD disk winds to hundred AU scale
  from the protostar}.
\newblock In \emph{European Physical Journal Web of Conferences}, volume~64 of
  \emph{European Physical Journal Web of Conferences}, page 05006, Jan. 2014.
\newblock \doi{10.1051/epjconf/20136405006}.

\bibitem[{Tabone} et~al.(2020){Tabone}, {Cabrit}, {Pineau des For{\^e}ts},
  {Ferreira}, {Gusdorf}, {Podio}, {Bianchi}, {Chapillon}, {Codella}, and
  {Gueth}]{Tab20}
B.~{Tabone}, S.~{Cabrit}, G.~{Pineau des For{\^e}ts}, and others.
\newblock {Constraining MHD disk winds with ALMA. Apparent rotation signatures
  and application to HH212}.
\newblock \emph{\aap}, 640:\penalty0 A82, Aug. 2020.
\newblock \doi{10.1051/0004-6361/201834377}.

\bibitem[{Xu} et~al.(2013){Xu}, {Li}, {Reid}, {Menten}, {Zheng}, {Brunthaler},
  {Moscadelli}, {Dame}, and {Zhang}]{Xu13}
Y.~{Xu}, J.~J. {Li}, M.~J. {Reid}, and others.
\newblock {On the Nature of the Local Spiral Arm of the Milky Way}.
\newblock \emph{\apj}, 769:\penalty0 15, May 2013.
\newblock \doi{10.1088/0004-637X/769/1/15}.

\bibitem[{Zhang} et~al.(2017){Zhang}, {Claus}, {Watson}, and {Moran}]{Zha17}
Q.~{Zhang}, B.~{Claus}, L.~{Watson}, and J.~{Moran}.
\newblock {Angular Momentum in Disk Wind Revealed in the Young Star MWC 349A}.
\newblock \emph{\apj}, 837\penalty0 (1):\penalty0 53, Mar. 2017.
\newblock \doi{10.3847/1538-4357/aa5ea9}.

\end{thebibliography}
%

\clearpage

\begin{table*}
\begin{center}
\begin{minipage}{\textwidth}
\caption{Parameters of the linear and sinusoidal fits}
\label{tab_fit}
\begin{tabular*}{\textwidth}{@{\extracolsep{\fill}}lccccc@{\extracolsep{\fill}}}
\hline\hline
& \multicolumn{2}{@{}c@{}}{Linear fit} & \multicolumn{3}{@{}c@{}}{Sinusoidal fit} \\ 
\cline{2-3} \cline{4-6} 
Stream & $\omega$ & $V_{\rm off}$ & $\mathfrak{R}$ & $f_z$ & $z_0$ \\
       & (km~s$^{-1}$~au$^{-1}$) & (km~s$^{-1}$) & (au) & (rad~au$^{-1}$) & (au) \\
\hline
SW  &  $1.15\pm0.12$ & $17.5\pm1.3$ & $21.6\pm1.0$ &  $0.0366\pm0.0012$ & $-16.5\pm1.9$\\
NE-1  &  $1.15\pm0.09$ & $-29.2\pm0.6$ & $36\pm5$ &  $0.0307\pm0.0008$ & $16.3\pm2.0$\\
NE-2  &  $0.64\pm0.27$ & $-10.9\pm3.0$ & $22.9\pm0.5$ &  $0.0386\pm0.0008$ & $47.8\pm1.2$\\
NE-3  &  $2.0\pm0.5^{a} $ & $-49\pm27$ & $17.2\pm0.3$ &  $0.0405\pm0.0023$ & $7.3\pm2.3$\\
\hline
\end{tabular*}
\tablefoot{
\\
Column~1 denotes the maser stream; 
Cols.~2~and~3 provide the values of \ $\omega$ \ and \  $V_{\rm off}$ \ from the linear fit of maser \Vlsr\ versus \ $R$; Cols.~4,~5~and~6 report the amplitude, the spatial frequency and the position of zero phase, respectively, of the sinusoidal fit of the maser coordinates \ $R$ \ versus \ $z$.
}   
\footnotetext[1]{The determination of this error, smaller than the value, 1.7~km~s$^{-1}$~au$^{-1}$, from the linear fit, is discussed in Appendix~\ref{met_NE3}.}
\end{minipage}
\end{center}
\end{table*}

\begin{table}[h]
\begin{minipage}{0.5\textwidth}
\caption{Estimate of the stream parameters}\label{tab_der}
\begin{tabular}{@{}lccc@{}}
\hline\hline
Stream & $V_z$ & $\omega_{\rm K}$ &  $\mathfrak{R}_{\rm K}$ \\
       & (km~s$^{-1}$) & (km~s$^{-1}$~au$^{-1}$)  & (au)  \\
\hline
SW    & $\ge$~48 &$\ge$~2.9 & $\le$~9 \\
NE-1  & $\ge$~46 & $\ge$~2.6  & $\le$~10 \\
NE-2  & $\ge$~9 & $\ge$~1.0  & $\le$~17 \\
NE-3  & $\ge$~85 & $\ge$~5.4  & $\le$~6 \\
\hline
\end{tabular}
\tablefoot{
\\
Column~1 indicates the maser stream; 
Col.~2 reports the estimated streaming velocity along the jet axis;
Cols.~3~and~4 give the estimate of the angular velocity and the radius, respectively, at the launch point.
}   
\end{minipage}
\end{table}

\clearpage

\begin{appendix}

\section{Observations} 
\label{met_obs}

We observed the  $6_{16} - 5_{23}$ H$_2$O maser transition (rest frequency 22.235079 GHz) towards \targ\ 
(tracking center: RA(J2000) = $21^{\rm
h} 9^{\rm m}$21\pss720 
and Dec(J2000) = +52\degree$22^{\prime}$37\pas08)
 with global VLBI for 24~hr, starting on 2020 October 27, 13:30~UT. The antennae involved were 16 antennae of the European VLBI network (EVN):
Yebes, Sardinia, Medicina, Jodrell\_Bank, Effelsberg, Onsala, Metsahovi, Torun, Svetloe, Badary, Zelenchukskaya, KVN\_Tamna, KVN\_Ulsan, KVN\_Yonsei, Urumqi, Tianma; plus the 10 antennae of the VLBA: Brewster, Fort Davis, Hancock, Kitt Peak, Los Alamos, Mauna Kea, North Liberty, Owens Valley, Pie Town, and Saint Croix. The observations were designed to achieve a relative and absolute position accuracy of \ $\sim$~0.01~mas and \ $\sim$~1~mas, respectively, and to reach a sensitivity in the maser line \ $\le 1$~m\Jyb. While the EVN antennae observed continuously the target, interleaving calibration scans \ $\approx$~8~min long every hour, the VLBA performed also phase-referencing observations (over 10~hr), alternating scans on the target and the phase-reference calibrator every 45~s. During the phase-reference session, the target and the calibrators were always observed by the VLBA simultaneously with the EVN to ensure global VLBI baselines. The fringe-finder and bandpass calibrators were: J2202$+$4216, 2007$+$777, 3C84 and 3C48; the phase-reference calibrators were \ 2116$+$543  \ and \ 2051$+$528, both within 2.5\degree \ from the target and with a correlated flux  of $\sim$~0.1~\Jyb \ at 22~GHz.

We recorded dual circular polarization through four adjacent bandwidths of 16~MHz, one of them centered at the maser \Vlsr \ of \ $-$6.4~\kms. The four 16~MHz bandwidths were used to increase the signal to noise ratio (SNR) of the weak continuum (phase-reference) calibrators. The
data were correlated with the SFXC correlator at the Joint Institute
for VLBI in Europe (JIVE, at Dwingeloo, the Netherlands) in two correlation passes,
using 1024 and 128 spectral channels to correlate the maser 16~MHz bandwidth
and the whole set of four 16~MHz bandwidths, respectively. 
The spectral resolution attained across the maser 16~MHz band was 0.21~\kms.
The correlator averaging time was 1~s. 

Data were reduced with the Astronomical Image Processing System (\textsc{AIPS}) package following the VLBI spectral line
procedures \cite{San10a}. 
The emission of an intense and compact maser channel was self-calibrated,
and the derived (amplitude \& phase) corrections were applied to all the maser channels
before imaging. To cover the whole maser emission, we produced images extending  \ 0\pas65  \ in both \ RA $\cos \delta$ \ and \ DEC, and 84~\kms \ in \Vlsr.
 Using natural weighting, the FWHM major and minor sizes of the beam are 
\ 0.7~mas and 0.3~mas, respectively, and the beam PA is \ $-$49\degree.
In channel maps with (relatively) weak signal, the 1$\sigma$ rms noise is 
0.7~mJy~beam$^{-1}$, close to the expected thermal noise.

Inverse phase-referencing \cite{San10a} produced good SNR ($\ge$~10)  images of the two phase-reference calibrators. Taking into account that the calibrators are relatively compact with size $\lesssim$~1~mas, and that the absolute position of the calibrators is known within a few 0.1~mas, we estimate that the error on the absolute position of the masers is \ $\lesssim$~0.5~mas.

Table~\ref{tab_mas} reports the parameters (intensity, \Vlsr, position)
of the 22~GHz water masers in \targ. Individual maser features are a collection of quasi-compact spots observed on contiguous channel maps and spatially overlapping (within their FWHM size). The spot positions are determined by fitting a two-dimensional elliptical Gaussian to their spatial emissions. The uncertainty of the spot position relative to the reference maser channel is the contribution of two terms: $\Delta \theta_{\rm spot} = \sqrt{\Delta \theta_{\rm fit}^2 + \Delta \theta_{\rm bandpass}^2}$. The first term depends on the SNR of the data, following \cite{Rei88}: \ $\Delta \theta_{\rm fit} = \theta_{\rm beam} \  / (2 \ {\rm SNR}) $, where \ $\theta_{\rm beam}$ \ is the resolution beam size, conservatively taken equal to the FWHM major beam size of 0.7~mas. The second  term depends on the accuracy of the bandpass calibration through the expression \cite{Zha17}: \ $\Delta \theta_{\rm bandpass} = \theta_{\rm beam} \  (\Delta \Psi / 360^{\circ})$, where \ $ \Delta \Psi$ \ in degrees is the phase stability across the observing band. In our case \ $ \Delta \Psi \lesssim 10^{\circ}$ \ and \   $\Delta \theta_{\rm bandpass} \lesssim 0.02$~mas \ becomes the dominant error term for spot intensity \ $\ge$~100~m\Jyb. The maser feature position (and corresponding error) is estimated from the 
error-weighted mean of the spot positions (and corresponding errors), and the feature \Vlsr \ from the intensity-weighted mean of the spots' \Vlsr. To be conservative, the uncertainty on the feature \Vlsr \ is taken equal  to \ 0.5~\kms, corresponding to the typical maser FWHM line width.

\longtab[3]{ 
\begin{longtable}{crrrr} 
\caption{\label{tab_mas} 22.2~GHz H$_2$O maser parameters for \targ}\\ 
\hline\hline
\multicolumn{1}{c}{Feature} &  \multicolumn{1}{r}{I$_{\rm peak}$} & \multicolumn{1}{r}{$V_{\rm LSR}$} & \multicolumn{1}{c}{$\Delta~x$} & \multicolumn{1}{c}{$\Delta~y$}  \\
\multicolumn{1}{c}{Number}  &  \multicolumn{1}{r}{(Jy beam$^{-1}$)} & \multicolumn{1}{r}{(km s$^{-1}$)} & \multicolumn{1}{c}{(mas)} & \multicolumn{1}{c}{(mas)}  \\
\hline
\endfirsthead
\caption{continued.}\\
\hline\hline
\multicolumn{1}{c}{Feature} &  \multicolumn{1}{r}{I$_{\rm peak}$} & \multicolumn{1}{r}{$V_{\rm LSR}$} & \multicolumn{1}{c}{$\Delta~x$} & \multicolumn{1}{c}{$\Delta~y$}  \\
\multicolumn{1}{c}{Number}  &  \multicolumn{1}{r}{(Jy beam$^{-1}$)} & \multicolumn{1}{r}{(km s$^{-1}$)} & \multicolumn{1}{c}{(mas)} & \multicolumn{1}{c}{(mas)}  \\
\hline
\endhead
\hline
\endfoot
\hline
\endlastfoot
    1 &    53.066 &   -21.6 &     0 \ \ \ \ \ \  &     0 \ \ \ \ \ \  \\
    2 &    14.326 &    -3.0 &    60.42$\pm$0.03 &    27.87$\pm$0.03  \\
    3 &    14.017 &   -24.0 &   -14.93$\pm$0.03 &    51.87$\pm$0.03  \\
    4 &    10.457 &   -28.5 &    -1.55$\pm$0.03 &    58.23$\pm$0.03  \\
    5 &     7.082 &     0.4 &   -42.37$\pm$0.03 &   -36.16$\pm$0.03  \\
    6 &     6.565 &    -1.0 &    56.89$\pm$0.03 &    18.87$\pm$0.03  \\
    7 &     6.502 &   -15.2 &    23.06$\pm$0.03 &    -4.24$\pm$0.03  \\
    8 &     5.543 &   -30.2 &    -1.29$\pm$0.03 &    59.10$\pm$0.03  \\
    9 &     4.904 &   -21.8 &     0.07$\pm$0.03 &     0.46$\pm$0.03  \\
   10 &     4.866 &   -15.1 &    33.59$\pm$0.03 &     3.34$\pm$0.03  \\
   11 &     4.812 &   -19.8 &    58.57$\pm$0.03 &    32.40$\pm$0.03  \\
   12 &     4.395 &   -24.8 &   -14.78$\pm$0.03 &    52.52$\pm$0.03  \\
   13 &     4.273 &     5.0 &   -43.75$\pm$0.03 &   -40.90$\pm$0.03  \\
   14 &     4.072 &    -3.2 &    60.79$\pm$0.03 &    28.54$\pm$0.03  \\
   15 &     3.556 &   -25.2 &    14.57$\pm$0.03 &    -3.37$\pm$0.03  \\
   16 &     3.397 &     1.5 &    61.14$\pm$0.03 &    31.34$\pm$0.03  \\
   17 &     3.387 &    -0.1 &   -41.84$\pm$0.03 &   -34.58$\pm$0.03  \\
   18 &     3.206 &   -15.4 &    33.76$\pm$0.03 &     3.88$\pm$0.03  \\
   19 &     3.177 &   -22.7 &   -11.28$\pm$0.03 &    22.53$\pm$0.03  \\
   20 &     2.485 &   -11.7 &    53.03$\pm$0.03 &    19.32$\pm$0.03  \\
   21 &     2.391 &   -17.3 &    29.96$\pm$0.03 &    -0.05$\pm$0.03  \\
   22 &     2.247 &    -2.1 &    59.06$\pm$0.03 &    22.92$\pm$0.03  \\
   23 &     2.154 &   -24.1 &   -15.03$\pm$0.03 &    51.42$\pm$0.03  \\
   24 &     2.072 &     4.3 &   -44.37$\pm$0.03 &   -39.19$\pm$0.03  \\
   25 &     2.062 &   -16.2 &    25.54$\pm$0.03 &    -3.08$\pm$0.03  \\
   26 &     1.952 &     4.5 &   -45.06$\pm$0.03 &   -41.80$\pm$0.03  \\
   27 &     1.926 &    -6.2 &    60.50$\pm$0.03 &    38.38$\pm$0.03  \\
   28 &     1.740 &   -27.4 &   -11.03$\pm$0.03 &    22.50$\pm$0.03  \\
   29 &     1.661 &   -37.8 &    55.01$\pm$0.03 &    33.58$\pm$0.03  \\
   30 &     1.631 &   -10.8 &    62.36$\pm$0.03 &    37.98$\pm$0.03  \\
   31 &     1.537 &     5.2 &   -43.54$\pm$0.03 &   -40.37$\pm$0.03  \\
   32 &     1.420 &   -27.5 &   121.15$\pm$0.03 &   120.88$\pm$0.03  \\
   33 &     1.137 &    -2.0 &    57.72$\pm$0.03 &    20.15$\pm$0.03  \\
   34 &     1.109 &    17.1 &   -11.32$\pm$0.03 &   -16.69$\pm$0.03  \\
   35 &     0.986 &    -1.6 &    58.65$\pm$0.03 &    22.43$\pm$0.03  \\
   36 &     0.954 &    -8.1 &    61.04$\pm$0.03 &    38.18$\pm$0.03  \\
   37 &     0.866 &   -13.6 &    21.30$\pm$0.03 &    -5.56$\pm$0.03  \\
   38 &     0.859 &    -9.5 &    44.93$\pm$0.03 &    10.49$\pm$0.03  \\
   39 &     0.774 &   -16.5 &    23.77$\pm$0.03 &    -3.55$\pm$0.03  \\
   40 &     0.756 &     0.4 &    59.72$\pm$0.03 &    27.76$\pm$0.03  \\
   41 &     0.745 &   -10.7 &    44.31$\pm$0.03 &    10.40$\pm$0.03  \\
   42 &     0.678 &   -22.7 &    14.29$\pm$0.03 &    -4.41$\pm$0.03  \\
   43 &     0.649 &    -8.9 &   120.25$\pm$0.03 &   100.23$\pm$0.03  \\
   44 &     0.554 &    -0.0 &   -42.04$\pm$0.03 &   -35.56$\pm$0.03  \\
   45 &     0.539 &   -30.3 &    -1.21$\pm$0.03 &    59.56$\pm$0.03  \\
   46 &     0.512 &     6.7 &   -57.34$\pm$0.03 &  -138.41$\pm$0.03  \\
   47 &     0.428 &    -8.9 &    44.04$\pm$0.03 &     9.57$\pm$0.03  \\
   48 &     0.375 &   -25.2 &    14.07$\pm$0.03 &    -3.40$\pm$0.03  \\
   49 &     0.374 &   -15.8 &    34.29$\pm$0.03 &     4.60$\pm$0.03  \\
   50 &     0.367 &   -23.3 &    14.55$\pm$0.03 &    -3.95$\pm$0.03  \\
   51 &     0.364 &    -4.3 &    60.56$\pm$0.03 &    30.64$\pm$0.03  \\
   52 &     0.358 &    -9.1 &    61.21$\pm$0.03 &    37.42$\pm$0.03  \\
   53 &     0.352 &   -18.1 &   118.48$\pm$0.03 &    99.46$\pm$0.03  \\
   54 &     0.339 &   -31.6 &    55.84$\pm$0.03 &    40.17$\pm$0.03  \\
   55 &     0.290 &    -8.5 &   121.39$\pm$0.03 &   100.59$\pm$0.03  \\
   56 &     0.270 &    -1.7 &    58.08$\pm$0.03 &    21.15$\pm$0.03  \\
   57 &     0.268 &    -3.5 &   138.43$\pm$0.03 &    90.24$\pm$0.03  \\
   58 &     0.266 &   -17.4 &    29.50$\pm$0.03 &    -0.42$\pm$0.03  \\
   59 &     0.259 &   -10.0 &    43.61$\pm$0.03 &     9.66$\pm$0.03  \\
   60 &     0.228 &   -25.8 &    -8.48$\pm$0.03 &    57.46$\pm$0.03  \\
   61 &     0.213 &   -30.6 &    55.99$\pm$0.03 &    39.54$\pm$0.03  \\
   62 &     0.200 &   -16.9 &    33.88$\pm$0.03 &     4.64$\pm$0.03  \\
   63 &     0.200 &   -12.2 &    21.80$\pm$0.03 &    -5.36$\pm$0.03  \\
   64 &     0.199 &   -10.1 &    62.03$\pm$0.03 &    38.42$\pm$0.03  \\
   65 &     0.195 &   -15.6 &    33.15$\pm$0.03 &     2.76$\pm$0.03  \\
   66 &     0.191 &     4.4 &   -45.46$\pm$0.03 &   -42.22$\pm$0.03  \\
   67 &     0.185 &   -16.5 &    24.85$\pm$0.03 &    -3.22$\pm$0.03  \\
   68 &     0.181 &    -8.5 &   124.31$\pm$0.03 &   100.94$\pm$0.03  \\
   69 &     0.178 &     0.5 &   -40.76$\pm$0.03 &   -32.20$\pm$0.03  \\
   70 &     0.170 &   -12.1 &    34.87$\pm$0.03 &     5.95$\pm$0.03  \\
   71 &     0.164 &    -1.2 &    60.11$\pm$0.03 &    34.24$\pm$0.03  \\
   72 &     0.161 &   -13.0 &    59.54$\pm$0.03 &    29.23$\pm$0.03  \\
   73 &     0.161 &    -1.9 &    61.56$\pm$0.03 &    32.23$\pm$0.03  \\
   74 &     0.161 &    -8.4 &   122.16$\pm$0.03 &   100.72$\pm$0.03  \\
   75 &     0.155 &   -19.9 &    58.58$\pm$0.03 &    31.90$\pm$0.03  \\
   76 &     0.155 &   -33.7 &    55.96$\pm$0.03 &    37.71$\pm$0.03  \\
   77 &     0.152 &     1.0 &    58.10$\pm$0.03 &    23.42$\pm$0.04  \\
   78 &     0.145 &     2.0 &   -42.80$\pm$0.03 &   -37.01$\pm$0.03  \\
   79 &     0.138 &   -27.5 &   -10.99$\pm$0.03 &    22.85$\pm$0.03  \\
   80 &     0.132 &   -26.0 &    30.09$\pm$0.03 &     0.66$\pm$0.03  \\
   81 &     0.127 &   -13.4 &    62.41$\pm$0.03 &    36.36$\pm$0.03  \\
   82 &     0.121 &     6.8 &    38.38$\pm$0.03 &  -105.32$\pm$0.03  \\
   83 &     0.107 &   -20.4 &    14.92$\pm$0.05 &    -5.16$\pm$0.05  \\
   84 &     0.105 &   -12.1 &    58.11$\pm$0.03 &    25.16$\pm$0.03  \\
   85 &     0.102 &    -1.0 &    61.01$\pm$0.03 &    30.58$\pm$0.04  \\
   86 &     0.101 &   -18.0 &    65.66$\pm$0.03 &    36.02$\pm$0.03  \\
   87 &     0.099 &   -18.6 &    -4.44$\pm$0.03 &    57.01$\pm$0.03  \\
   88 &     0.097 &   -23.5 &   -15.53$\pm$0.03 &    49.39$\pm$0.03  \\
   89 &     0.096 &     1.6 &    57.02$\pm$0.03 &    21.67$\pm$0.03  \\
   90 &     0.094 &    -1.2 &   -44.13$\pm$0.03 &   -38.22$\pm$0.03  \\
   91 &     0.094 &     2.9 &    60.86$\pm$0.03 &    30.86$\pm$0.03  \\
   92 &     0.093 &     5.7 &   -42.83$\pm$0.03 &   -39.80$\pm$0.03  \\
   93 &     0.091 &   -34.2 &   -12.84$\pm$0.03 &    51.09$\pm$0.03  \\
   94 &     0.074 &   -26.3 &   -11.14$\pm$0.03 &    22.22$\pm$0.03  \\
   95 &     0.069 &   -18.4 &    30.26$\pm$0.03 &     0.51$\pm$0.03  \\
   96 &     0.069 &    -8.3 &   127.68$\pm$0.03 &   100.86$\pm$0.03  \\
   97 &     0.067 &    -0.2 &   -40.55$\pm$0.05 &   -31.83$\pm$0.04  \\
   98 &     0.065 &     0.0 &    58.01$\pm$0.04 &    21.83$\pm$0.05  \\
   99 &     0.065 &   -18.2 &   -12.24$\pm$0.03 &    22.63$\pm$0.03  \\
  100 &     0.064 &   -17.9 &   -20.70$\pm$0.03 &    26.63$\pm$0.03  \\
  101 &     0.064 &   -19.8 &   -22.59$\pm$0.03 &    32.39$\pm$0.03  \\
  102 &     0.063 &   -10.9 &    62.10$\pm$0.03 &    37.24$\pm$0.03  \\
  103 &     0.062 &   -10.6 &    35.68$\pm$0.03 &     8.84$\pm$0.03  \\
  104 &     0.061 &    17.0 &   -11.54$\pm$0.03 &   -16.91$\pm$0.03  \\
  105 &     0.060 &   -14.3 &    41.60$\pm$0.03 &    10.24$\pm$0.03  \\
  106 &     0.056 &     9.1 &   -45.72$\pm$0.03 &   -44.04$\pm$0.03  \\
  107 &     0.051 &    -0.9 &   -21.86$\pm$0.05 &   -17.36$\pm$0.04  \\
  108 &     0.049 &    -9.9 &    37.06$\pm$0.03 &     7.68$\pm$0.03  \\
  109 &     0.048 &   -14.7 &    59.47$\pm$0.03 &    28.61$\pm$0.03  \\
  110 &     0.048 &     2.7 &    58.82$\pm$0.03 &    25.55$\pm$0.03  \\
  111 &     0.047 &   -12.2 &    59.77$\pm$0.03 &    29.64$\pm$0.03  \\
  112 &     0.046 &   -12.5 &    43.30$\pm$0.03 &    11.05$\pm$0.03  \\
  113 &     0.046 &    -8.3 &   -41.69$\pm$0.03 &   -23.95$\pm$0.03  \\
  114 &     0.044 &   -12.5 &    62.24$\pm$0.03 &    35.91$\pm$0.03  \\
  115 &     0.044 &    -1.1 &   -23.09$\pm$0.06 &   -18.46$\pm$0.05  \\
  116 &     0.043 &   -16.0 &    59.12$\pm$0.04 &    27.70$\pm$0.04  \\
  117 &     0.043 &   -33.2 &   -13.02$\pm$0.03 &    50.88$\pm$0.03  \\
  118 &     0.041 &    11.6 &    12.77$\pm$0.03 &   -29.04$\pm$0.03  \\
  119 &     0.040 &    10.5 &    12.91$\pm$0.03 &   -27.04$\pm$0.03  \\
  120 &     0.039 &   -11.5 &    44.63$\pm$0.03 &    10.64$\pm$0.03  \\
  121 &     0.039 &    -7.7 &    36.05$\pm$0.03 &     7.54$\pm$0.03  \\
  122 &     0.036 &    -5.9 &    60.39$\pm$0.03 &    37.90$\pm$0.03  \\
  123 &     0.036 &    11.0 &   -50.13$\pm$0.03 &  -135.02$\pm$0.03  \\
  124 &     0.033 &   -27.4 &   121.11$\pm$0.04 &   120.44$\pm$0.03  \\
  125 &     0.032 &   -18.4 &   -21.05$\pm$0.03 &    27.82$\pm$0.03  \\
  126 &     0.032 &     3.6 &   -43.27$\pm$0.03 &   -38.14$\pm$0.04  \\
  127 &     0.032 &   -31.6 &    56.14$\pm$0.03 &    40.39$\pm$0.03  \\
  128 &     0.030 &   -29.4 &    54.79$\pm$0.03 &    35.16$\pm$0.03  \\
  129 &     0.029 &   -10.8 &    46.33$\pm$0.04 &    11.49$\pm$0.03  \\
  130 &     0.029 &   -20.1 &   -22.18$\pm$0.04 &    37.50$\pm$0.04  \\
  131 &     0.027 &     9.7 &   -59.68$\pm$0.03 &  -137.27$\pm$0.03  \\
  132 &     0.025 &    -5.8 &   110.29$\pm$0.04 &    56.94$\pm$0.04  \\
  133 &     0.022 &   -14.1 &    34.57$\pm$0.04 &     5.08$\pm$0.04  \\
  134 &     0.021 &     5.9 &   -42.25$\pm$0.03 &   -39.12$\pm$0.03  \\
  135 &     0.020 &   -29.4 &    56.24$\pm$0.04 &    35.30$\pm$0.04  \\
  136 &     0.019 &    10.9 &    13.38$\pm$0.04 &   -27.78$\pm$0.04  \\
  137 &     0.019 &   -18.3 &    24.24$\pm$0.04 &    -3.28$\pm$0.04  \\
  138 &     0.018 &    10.8 &   -59.98$\pm$0.04 &  -136.68$\pm$0.04  \\
  139 &     0.017 &    29.3 &   -59.75$\pm$0.03 &  -120.52$\pm$0.03  \\
  140 &     0.017 &    -5.8 &   108.81$\pm$0.05 &    57.04$\pm$0.05  \\
  141 &     0.017 &   -18.3 &   118.43$\pm$0.05 &    98.90$\pm$0.04  \\
  142 &     0.016 &   -31.0 &    64.36$\pm$0.04 &    45.69$\pm$0.05  \\
  143 &     0.015 &    -7.6 &   134.19$\pm$0.04 &    91.17$\pm$0.04  \\
  144 &     0.015 &    19.2 &   -48.27$\pm$0.04 &   -55.22$\pm$0.04  \\
  145 &     0.013 &     7.3 &    34.56$\pm$0.04 &  -116.70$\pm$0.04  \\
  146 &     0.013 &   -12.7 &    39.14$\pm$0.05 &     9.70$\pm$0.05  \\
  147 &     0.012 &    -8.8 &   119.61$\pm$0.05 &   100.15$\pm$0.05  \\
  148 &     0.011 &    13.5 &    11.24$\pm$0.04 &   -30.82$\pm$0.04  \\
  149 &     0.011 &   -33.5 &   -13.27$\pm$0.04 &    50.41$\pm$0.04  \\
  150 &     0.011 &    10.4 &    12.78$\pm$0.04 &   -26.19$\pm$0.04  \\
  151 &     0.011 &    -5.5 &   111.16$\pm$0.05 &    56.61$\pm$0.05  \\
  152 &     0.011 &    19.0 &   -60.91$\pm$0.04 &  -124.10$\pm$0.04  \\
  153 &     0.010 &    13.4 &    12.64$\pm$0.04 &   -32.15$\pm$0.04  \\
  154 &     0.010 &    -5.3 &   140.28$\pm$0.05 &    33.19$\pm$0.05  \\
  155 &     0.010 &   -31.7 &     1.13$\pm$0.04 &    61.26$\pm$0.04  \\
  156 &     0.009 &    17.9 &   -63.61$\pm$0.05 &  -124.63$\pm$0.05  \\
  157 &     0.009 &    -7.6 &   123.27$\pm$0.06 &   101.80$\pm$0.05  \\
  158 &     0.009 &   -18.7 &   -21.77$\pm$0.06 &    29.92$\pm$0.05  \\
  159 &     0.008 &     8.6 &    33.57$\pm$0.05 &  -119.04$\pm$0.05  \\
  160 &     0.008 &   -33.0 &   -12.72$\pm$0.05 &    51.44$\pm$0.05  \\
  161 &     0.008 &    -5.4 &    83.67$\pm$0.06 &    45.45$\pm$0.07  \\
  162 &     0.008 &    -9.1 &    98.09$\pm$0.09 &    68.97$\pm$0.08  \\
  163 &     0.007 &   -39.6 &   140.57$\pm$0.06 &   213.71$\pm$0.06  \\
  164 &     0.007 &   -37.1 &   139.82$\pm$0.06 &   214.08$\pm$0.06  \\
  165 &     0.006 &    11.9 &    13.04$\pm$0.06 &   -29.94$\pm$0.05  \\
  166 &     0.006 &     8.6 &    40.93$\pm$0.06 &   -85.43$\pm$0.07  \\
  167 &     0.006 &    10.3 &   -62.68$\pm$0.08 &   -66.21$\pm$0.06  \\
  168 &     0.005 &    12.5 &   -45.00$\pm$0.06 &   -44.03$\pm$0.05  \\
\end{longtable}
\tablefoot{
\\
Column~1 gives the feature label number; 
Cols.~2~and~3 provide the intensity of the strongest spot
and the intensity-weighted spots' \Vlsr, respectively; Cols.~4~and~5 give the position offsets (with
the associated errors) towards East and North, respectively, relative to feature~\#~1. \\
The absolute position of feature~\#~1 is:
R.A.~(J2000) = 21$^{\rm h}$ 9$^{\rm m}$ 21\fs7099,  
Dec~(J2000) = 52\degree\ 22$^{\prime}$ 37\farcs001. 
The absolute position is evaluated at the observing epoch: October 27, 2020. 
}   
}

\section{Resolving the NE emission into three distinct streams} 
\label{met_NE1}

The masers in the NE emission are separated into three distinct streams identified with different symbols: dots, triangles and squares (see Fig.~\ref{NE_spi}a). Each stream traces a sinusoid in the plane of the sky (see Fig.~\ref{NE_spi}c), which is the signature for a spiral motion. The identification of the maser features belonging to each of the three sinusoids can be done with good confidence. First, we note that the  masers at elevation \ $z \ge 90$~au \ can be unambiguously divided into two streams, dots and triangles, on the basis of the very different \Vlsr \ of nearby emissions. Both dots and triangles at \ $z \ge 90$~au \  have \ $R$ \ decreasing with \ $z$, suggesting that they could trace the descending portion of a sinusoid, and one can argue that the corresponding ascending portion would be traced by masers with 
 \ $z < 90$~au. Next, we note that the group of masers with \ 30~au $\le z \le$~60~au \ draw an arc, and, since they cannot be part of a sinusoid together with other masers at larger $R$, they have to trace a third sinusoid with maximum radius close to the radius, $R_{\rm apex} = 17.2$~au, of the apex of the arc. We have also verified that the blue-shifted cluster of masers located at elevation \ $z \approx$~20~au (inside the dashed rectangle of Fig.~\ref{NE_spi}a) cannot be adjusted within this third sinusoid. We have fitted the expression:
\begin{linenomath*}
\begin{align}
R = (R_{\rm apex}-C) +  C \ \sin(f_z (z-z_0)) \label{eq_apex} 
\end{align}
\end{linenomath*}
with the maximum radius fixed to \ $R_{\rm apex}$ \ and the zero level given by \ $ R_{\rm apex}-C$, with \ $C$ \  free parameter.
When the masers at elevation \ $z \approx$~20~au \ and those drawing the arc at \ 30~au $\le z \le$~60~au \ are considered together, the fit of Eq.~\ref{eq_apex} results into an implausible, very negative zero level at \ $\approx -$60~au. Thus, after putting all the masers with \ 30~au $\le z \le$~60~au (and only them) into a third spiral motion, it remains to  deal only with the masers in the two separated ranges  \ 20~au $\le z <$~30~au \ and \ 60~au $\le z <$~90~au. The sinusoidal fit of \ $R$ versus $z$ \ of the triangles at \ $z \ge 90$~au \ indicates that the ascending portion of the sinusoid has to extend from \ $\approx$~50~au to \ $\approx$~90~au, and, over that range of elevations, intercepts all the observed masers. Finally, the remaining masers at  \ 20~au $\le z <$~30~au \  trace the ascending portion of the sinusoid identified with dots in  Fig.~\ref{NE_spi}a.

\begin{figure*}
\centering
\includegraphics[width=0.7\textwidth]{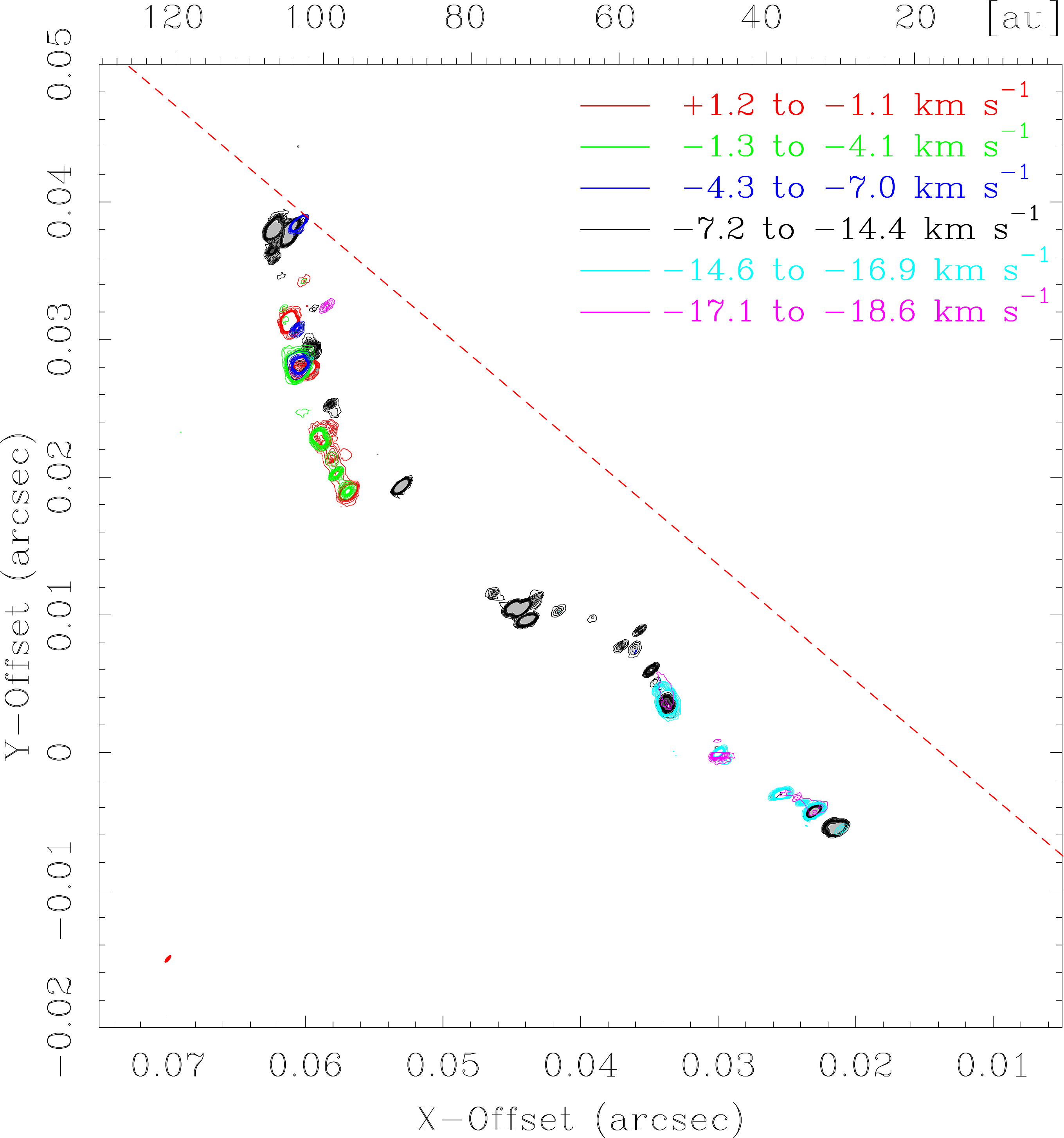}
\caption{Integrated emission map of the NE maser stream (see Fig.~\ref{NE_spi}a). 
Emission summed over the whole velocity range \ [$-$19, $+$1]~\kms\ by selecting groups of channels (top right) 
with similar noise thresholds. Colored contours correspond to a given velocity interval and increase at steps of 
5 times the noise thresholds of each group (corresponding to \ $\approx$~10\,$\sigma$). 
Angular offsets correspond to Table~\ref{tab_mas} and the linear scale is drawn in the top axis. The dashed red line denotes the jet axis and the synthesized VLBI beam is plotted in the bottom left corner.}
\label{combo}
\end{figure*}

Figure~\ref{combo} evidences the filamentary morphology of the maser emission in the NE region. First, we selected six velocity ranges (see Fig.~\ref{combo}) of channel maps with similar noise thresholds. Then, we summed the emission of each group of channels, 
blanking pixels with emission below a typical threshold of \ $\approx$~10\,$\sigma$, and overplotted the six maps together. This method allows to appreciate the overall spatial structure traced by maser emission with very different brightness levels.

\section{The angular velocity of the NE-3 stream} 
\label{met_NE3}

\begin{figure*}%
\centering
\includegraphics[width=0.7\textwidth]{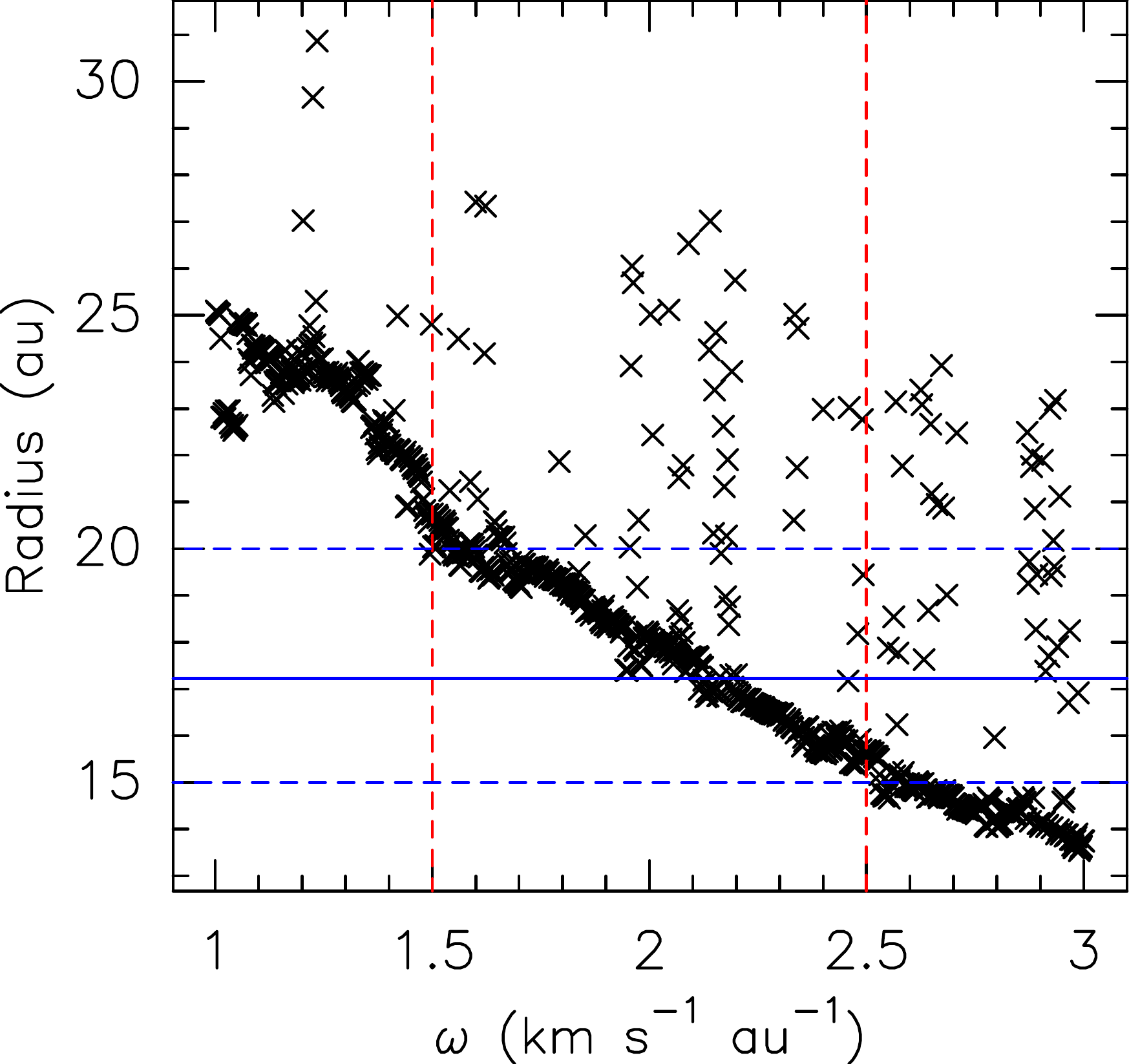}
\caption{Constraining \ $\omega$ \ for the NE-3 stream. Plot of the rotation radius estimated from the ratio \ $ \mathfrak{C} \ / \omega$ \ versus \ $\omega$, where
 \ $\mathfrak{C}$ \ is the amplitude of the sinusoidal fit of the rotational velocity, $\omega \ R $,  versus $z$ 
 (see Appendix~\ref{met_NE3}). The blue continuous and dashed lines mark the observed value of the rotation radius and the corresponding range of uncertainty, respectively; the red dashed lines delimit the range of \ $\omega$ \ for which the fitted radius is consistent with the observations.}
 \label{NE-3_bf}
\end{figure*}

The NE-3 stream is sampled by the masers only close to the positions where it attains the maximum radius. Because of the small radial range, 15~au $ \lesssim R \lesssim $~17~au, the linear fit of \ \Vlsr \  versus $R$ \ constrains very little the angular velocity, $\omega = 2.04\pm1.72$~\kmsau. However, since \ $R$, and the rotational velocity \ $\omega \ R $, are sinusoidal functions of the elevation \ $z$, we can use the results of the sinusoidal fits to reduce the error in $\omega $. For plausible values of \ $\omega$ \ in the range \ 0--4, we have performed the sinusoidal fit of the rotational velocity, and estimated the radius from the ratio of the fitted amplitude, $\mathfrak{C}$, and \ $\omega$.  For a given value of \ $\omega$, we have fitted the expression:
\begin{linenomath*}
\begin{align}
\omega \ R = (\omega \ R_{\rm apex}-\mathfrak{C}) +  \mathfrak{C} \ \sin(f_z (z-z_0)) \label{eq_apex2} 
\end{align}
\end{linenomath*}
fixing the maximum velocity  to \ $\omega \ R_{\rm apex}$ \ and fitting the zero level \ $\omega \ R_{\rm apex}-\mathfrak{C}$, with the amplitude \ $\mathfrak{C}$ \ as a free parameter. Fig.~\ref{NE-3_bf} shows that, apart a minor fraction of scattered results, the radius estimated from the sinusoidal fit decreases regularly with \ $\omega$, and attains a value consistent with the observations, $17.5\pm2.5$~au, for \ 
$\omega = 2.0\pm0.5$~\kmsau, thus effectively reducing the uncertainty to \ 0.5~\kmsau.

\section{Velocity scatter} 
\label{met_res}

Figs.~\ref{SW_spi}b~and~\ref{NE_spi}b show that the maser \Vlsr\ are linearly correlated with \ $R$ \ in the SW and NE-1 streams. 
For the masers belonging to the NE-2 stream, the measurement scatter from the linear fit of \Vlsr \ versus  \ $R$ \ is considerable (with large fit errors, see Table~\ref{tab_fit}), while, in the case of the NE-3 stream, the range in \ $R$ \ is too small (2~au only) to constrain the parameters of the linear fit. The noticeable deviation of \Vlsr \ from the linear fits seems difficult to conciliate with the accuracy with which the maser positions reproduce the sinusoidal patterns (see Figs.~\ref{SW_spi}c~and~\ref{NE_spi}c). The average scatters of maser \Vlsr\ and positions (along the jet axis from the fitted sinusoid) are  \ 2.4~\kms\ and \ 1.9~au, 3.1~\kms \ and \ 3.8~au, 6.9~\kms \ and \ 2.1~au, and 2.7~\kms \ and \ 1.7~au, for the SW, NE-1, NE-2 and NE-3 streams, respectively. Assuming that the \Vlsr\ scatter reflects mainly the variation of the streaming velocity \ $V_z$, since the jet axis is within \ 30$^{\circ}$ \ from the plane of the sky, the change in \ $V_z$ \ has to be at least twice that observed in \Vlsr. Then, using the above values, the ratios of the corresponding scatters in position and \ $V_z$ \ give upper limits for the maser traveling times of \ $\le$~1--3~yr. These traveling times, for maximum (absolute) elevations of the maser streams of \ 100--130~au (see Figs.~\ref{SW_spi}~and~\ref{NE_spi}), imply \ $V_z \ge$~200--500~\kms, which appears to be too large with respect to the observed maser \Vlsr. In the following, we investigate the physical reason for the observation of well defined sinusoidal patterns despite the co-occurrence of significant velocity scatters.  

In a stationary, axisymmetric MHD flow, the two fundamental equations of motions \cite{Pel92,Pud07,Pud19} linking velocity and magnetic field along a field line are:
\begin{linenomath*}
\begin{align}
\rho V_{\rm p} &= k B_{\rm p}   \label{eq_vbp} \\
B_{\phi} &= \frac{\rho}{k} \ (V_{\phi} - \omega_{\rm K} \mathfrak{R})  \label{eq_vbt}
\end{align}
\end{linenomath*}
where \ $V_{\rm p}$ \ and \ $V_{\phi}$, and \ $B_{\rm p}$ \ and \ $B_{\phi}$ \ are the poloidal and toroidal components of the velocity and magnetic field, respectively, $\rho$ \ is the gas mass volume density,  $\omega_{\rm K}$ \ is the Keplerian angular velocity at the launch point,  $\mathfrak{R}$ is the rotation radius, and  $k$ \ is the "mass load" of the wind, expressing the fixed ratio of mass and magnetic fluxes along a given magnetic field line. 
Since \ $ V_{\phi} - \omega_{\rm K} \ \mathfrak{R}$ \ is the toroidal  velocity in the reference frame co-rotating with the launch point, Eqs.~\ref{eq_vbp}~and~\ref{eq_vbt} lead to the well-known result that the velocity and magnetic field vectors are always paralell in the co-rotating reference frame.
Writing \ $ V_{\phi} = \omega \ \mathfrak{R} $, where \ $\omega$ \ is the angular velocity of the trajectory at radius \ $\mathfrak{R}$, the two equations above can be combined into:
\begin{linenomath*}
\begin{align}
\frac{B_{\phi}}{B_{\rm p}} =   \frac{(\omega - \omega_{\rm K})}{V_{\rm p}} \ \mathfrak{R} =  \frac{\omega_{\rm e}}{V_{\rm p}} \ \mathfrak{R} \label{eq_bpbt}
\end{align}
\end{linenomath*}
 where  we have used the definition of \ $\omega_{\rm e}$ \ in Eq.~\ref{eq_we}. We can define the magnetic field helix angle 
 \ $\alpha_B = \arctan(B_{\phi} / B_{\rm p}) $, which is the angle with which a helical field line winds around the jet axis.

We have seen that the observation of well defined sinusoids in the plane of the sky requires that the rotation radius \ $\mathfrak{R}$, and the ratio of the effective angular velocity \ $\omega_{\rm e}$ \ on the streaming velocity \ $V_z$ \ keep constant (see Eq.~\ref{eq_fz2}). Since the poloidal velocity \ $ V_{\rm p}$ \ is equal to \ $V_z$ \ if \ $\mathfrak{R}$ \ is constant, Eq.~\ref{eq_bpbt} implies that \ $ \alpha_B $ \ does not vary along each of the observed maser streams.
However, the reverse argument holds too. If the magnetic field is sufficiently stable and has a constant helix angle (that is, a helical configuration), the motion along the field line (in the co-rotating reference frame) requires a constant value of the ratio \ $ \omega_{\rm e} / V_z$. That preserves the maser sinusoidal pattern
despite the concomitant presence of a relatively large velocity scatter. From Eqs.~\ref{eq_bpbt}~and~\ref{eq_fz2}, with \ $V_{\rm p} = V_z$, we have:
\begin{linenomath*}
\begin{align}
\frac{|B_{\phi}|}{B_{\rm p}} = f_z \ \mathfrak{R}
\end{align}
\end{linenomath*}
Using the values reported in Table~\ref{tab_fit}, we derive:  $f_z \ \mathfrak{R} =$~$0.79\pm0.06$, $1.1\pm0.2$, $0.88\pm0.04$, and \ $0.70\pm0.05$ \ for the SW, NE-1, NE-2 and NE-3 streams, respectively. 
From Eq.~\ref{eq_vbt}, $B_{\phi} \approx 0$ \ within the Alfv\'en point when the flow co-rotates with the disk and \ $ V_{\phi} \approx \omega_{\rm K} \ \mathfrak{R} $. Beyond the Alfv\'en point, the inertia of the matter in the flow forces the magnetic field to fall behind the rotation of the disk, which causes \ $|B_{\phi}|$ \ to increase. The derived values of the ratio \ $ |B_{\phi}| /  B_{\rm p} $ in the range \ 0.7--1.1 agree with the water masers tracing a region of the spiraling trajectory in between the Alfv\'en point and the fast magneto-sonic point, beyond which the toroidal magnetic field component should become the predominant component \cite{Pud07,Sta14}. 

\begin{figure*}%
\centering
\includegraphics[width=0.5\textwidth]{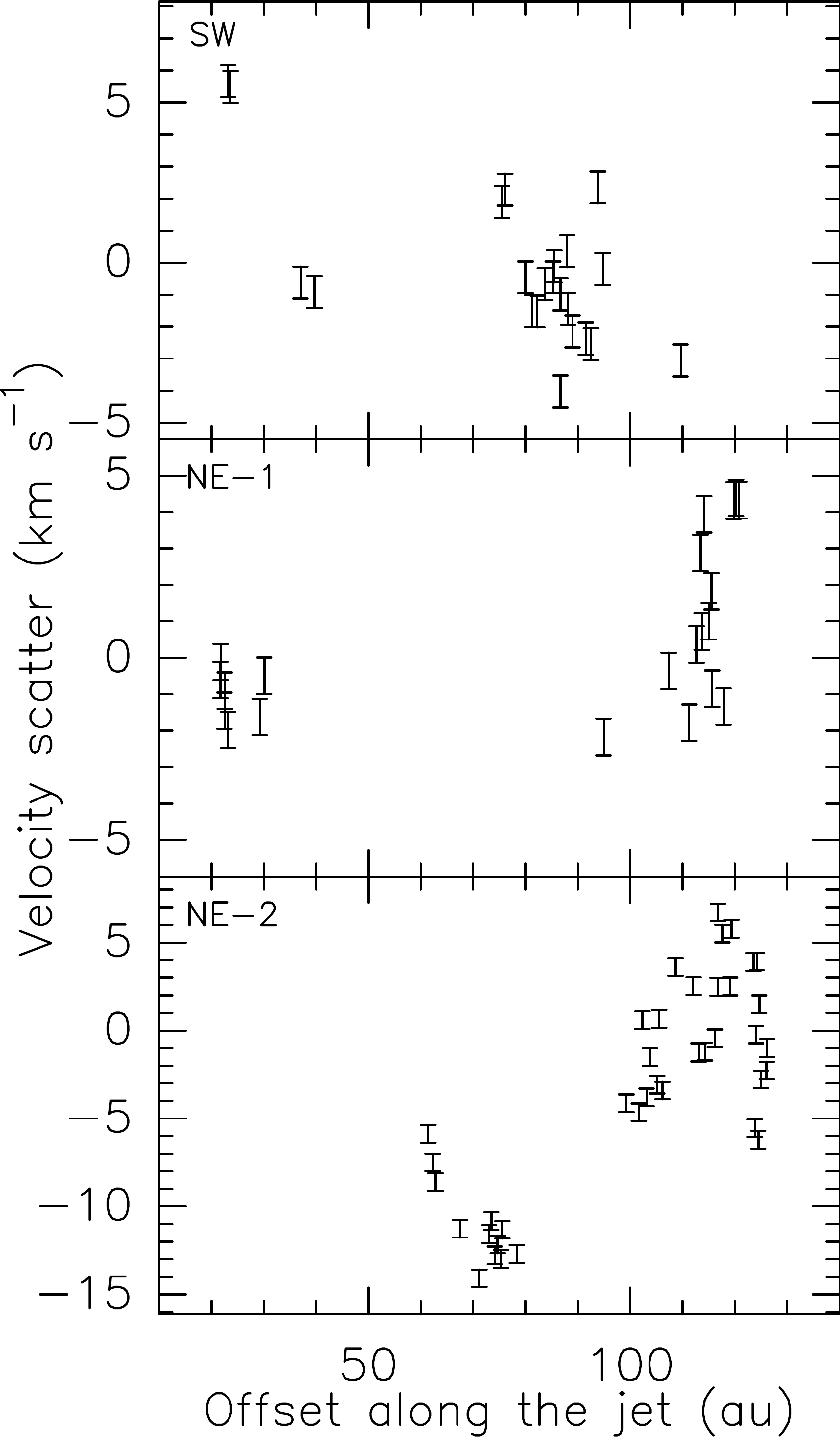}
\caption{Maser velocity scatter. The residual velocity from the linear fit of maser \Vlsr\ versus \ $R$ (see Figs.~\ref{SW_spi}b~and~\ref{NE_spi}b) is plotted versus \ $|z|$ \ for the masers in the SW (upper panel), NE-1 (central panel), and NE-2 (lower panel) streams.}
\label{resid}
\end{figure*}

Finally, we discuss briefly possible causes for the observed velocity scatter.
For the SW, NE-1, and NE-2 spiral motions, the residual of the linear fit of \Vlsr\ versus \ $R$ \ is plotted versus \ $z$ \ in Fig.~\ref{resid}. These plots indicate that the residual velocity is not correlated with position, contrarily to what it would be expected if the velocity scatter from the linear fit was caused by the smooth change with \ $z$ \ of \ the characteristic velocities, $\omega$ \ and \ $V_z$ (and consequently \ $V_{\rm off}$), of the spiraling trajectory. 
The residual velocities cannot be dominated by turbulence, too, since the latter would produce an independent variation of the toroidal and poloidal velocity components, which, as discussed above, is not consistent with the observation of the maser sinusoidal patterns. Moreover, it would be difficult to explain the observed different degree of turbulence between the NE-1 and NE-2 streams (see Fig.~\ref{resid}) over the same region of the sky.

The irregular change of the velocities can be naturally linked to our flow tracers, the water masers, which are excited in internal shocks of the flow emerging from perturbed locations of the disk. Eq.~\ref{eq_vbp}~and~\ref{eq_vbt} show that the flow velocity depends critically on the mass load \ $k$, which, on its turn, depends on the physical conditions, including dissipative processes, near the disk surface.
If the physical conditions of the perturbed launch points varied erratically on time scales so short as months, that would qualitatively explain the observed scatter of maser velocities. Emission variability over time scales of months is observed and theoretically expected in accretion bursts from high-mass YSOs \cite{Car17,Hun17,Oli20}, in such cases where an extended portion of the accretion disk gets perturbed.


\section{Water masers: shock type and corresponding flow kinematics}
\label{met_shock}
%

We propose that the water masers observed in \targ\ inside the NE, SW and N regions arise in internal shocks along streamlines of the YSO's disk wind. If observed at sufficiently high angular resolution, radio jets are often resolved into chains of compact emission centers (named ``knots''), which are interpreted in terms of internal shocks owing to fluctuation in mass or velocity ejection\cite{Ang18}. Since the jets are the innermost (most collimated and fastest) portion of a disk wind, it is possible that the structure of more external streamlines, where the gas is weakly ionized and mainly molecular, is characterized by internal shocks, too.  If the fluctuation responsible for the shocks is relatively small compared to the (time average) flow properties, the internal shocks are weak C-shocks and the shock (and maser) velocity is close to the flow velocity. In fact, the proper motions of radio knots are considered a reliable measurement of the jet speed\cite{Ang18}. In \targ, the regular (arc-like) spatial and \Vlsr \ distribution of the water masers indicate that the masers co-move with the flowing gas.

Models of water maser excitation predict them to arise in both strong dissociative J\cite{Eli89,Hol13}- and 
weak non-dissociative C\cite{Kau96}-shocks. While in J-shocks the masers form in the cooling post-shock gas at \ $T_{\rm kin} \approx 400$~K  following the chemical re-formation of water, in C-shocks the masing molecules do not dissociate and reach a significantly higher temperature, $T_{\rm kin} \approx 1000-3000$~K. The presence of shocks is naturally expected in correspondence of the fast protostellar outflows, whose association with the 22 GHz water masers is well proved observationally\cite{Mos20}. The location of the 22~GHz masers, often found at the wall of the jet cavity or at the terminal bow-shock of the jet, and their speed (mainly $\le$ 30~\kms)\cite{Mos21} significantly smaller that that of the jet (a few 100~\kms)\cite{Ang18}, indicate that the large majority of the 22~GHz water masers arise in strong external (lateral or terminal) J-shocks of the fast flow against the dense ambient material. While the velocity of the internal shocks is close to the flow speed, external shocks travel at a reduced speed because of the density contrast between the lighter impinging flow and the denser surrounding material.

As exemplified by our observations in \targ, the origin of the 22~GHz masers in internal shocks of the flow make them a formidable tool to trace individual streams of gas of the disk wind. In the following, we show that the alternative interpretation for these masers as external shocks of a wide-angle wind or a jet impacting against the surrounding medium is much less plausible. Looking at Fig.~\ref{glo}a, the water masers could be found at the wall of a wind cavity, if the axis of this putative wind was oriented at PA $\approx$~10\degree--20\degree \ and its opening angle was $\approx$~30\degree--40\degree. It is very difficult to conciliate this wide-angle wind with the jet geometry and precession rate estimated from previous (2010--2013) JVLA and VLBA observations (see Fig.~\ref{NV}). The elongation (PA = 56\degree$\pm$12\degree) of the slightly resolved (size $\approx$~150~au) JVLA continuum at 1.3~cm, the collimated (PA = 49\degree$\pm$18\degree) proper motions of the water masers near (100--200~au) the YSO, and the direction (PA $\approx$~44\degree) of the non-thermal radio jet traced by the extended (size $\approx$~1000~au) JVLA continuum at 5~cm, consistently indicate the presence of a jet oriented at \ PA $\approx$~50\degree. A distinct signature of precession\cite{Mos21} was the finding of a spur in the 5~cm continuum \ $\approx$~7000~au NE from the YSO along the same direction, PA = 60\degree, of the red-shifted lobe of the SO outflow observed with NOEMA. Clearly, a precession rate of only $\approx$~(60\degree$-$50\degree) = 10\degree \ over a length scale of 7000~au cannot account for a change in orientation between the putative wind and the JVLA/VLBA jet of \ $\ge$~(50\degree$-$20\degree) = 30\degree  \ over a length scale $\le$~100~au and a timescale \ $\le$10~yr. On the other hand, if this inner wind and the jet co-existed, it would be even more difficult to explain an abrupt transition in orientation and opening angle from the former to the latter across a few 10~au (see Fig.~\ref{glo}a). The interpretation of the water masers in \targ\ as external shocks at the cavity wall of a wide-angle wind faces also the difficulty to account for the huge maser \Vlsr \ gradients, $\ge$ 40~\kms \ over $\le$~10~au, transversal to the NE stream at elevation \ $z \ge 90$~au 
(see Fig.~\ref{NE_spi}a). If the masers, instead of delineating two streams at different velocities, traced shock fronts (seen close to edge-on) at the wind wall, it would be very difficult to explain such a large velocity difference for almost overlapping emissions. 
 
We conclude this discussion on alternatives to the disk wind interpretation showing that it is improbable that we are observing multiple outflows from a binary system. Inspecting Fig.~\ref{NV}, our previous NOEMA/JVLA/VLBA observations consistently indicate a single disk/jet system on scales of 100--500~au. The NOEMA 1.37~mm and the JVLA 1.3~cm continuum emissions, tracing the hosting molecular core and YSO's ionization, respectively, are compact and well aligned in position. The rotating disk revealed by the \Vlsr\ gradient of high-density molecular (CH$_3$CN and HC$_3$N) lines is centered on the peak of the 1.3~cm continuum. The agreement in direction between the elongated (double-lobe) JVLA 5~cm continuum and the collimated water maser proper motions points to a single jet emerging from the YSO. Considering now the new, global VLBI observations, on scales of 1--100~au, Fig.~\ref{glo}a shows that the spatial distribution of the water masers is about symmetrical with respect to the YSO position. This further supports the interpretation that a single YSO is responsible for the excitation and motion of the water masers.

\section{Constraining the Alfv\'en point of the NE-1 stream} 
\label{met_wk}

\begin{figure*}%
\centering
\includegraphics[width=0.5\textwidth]{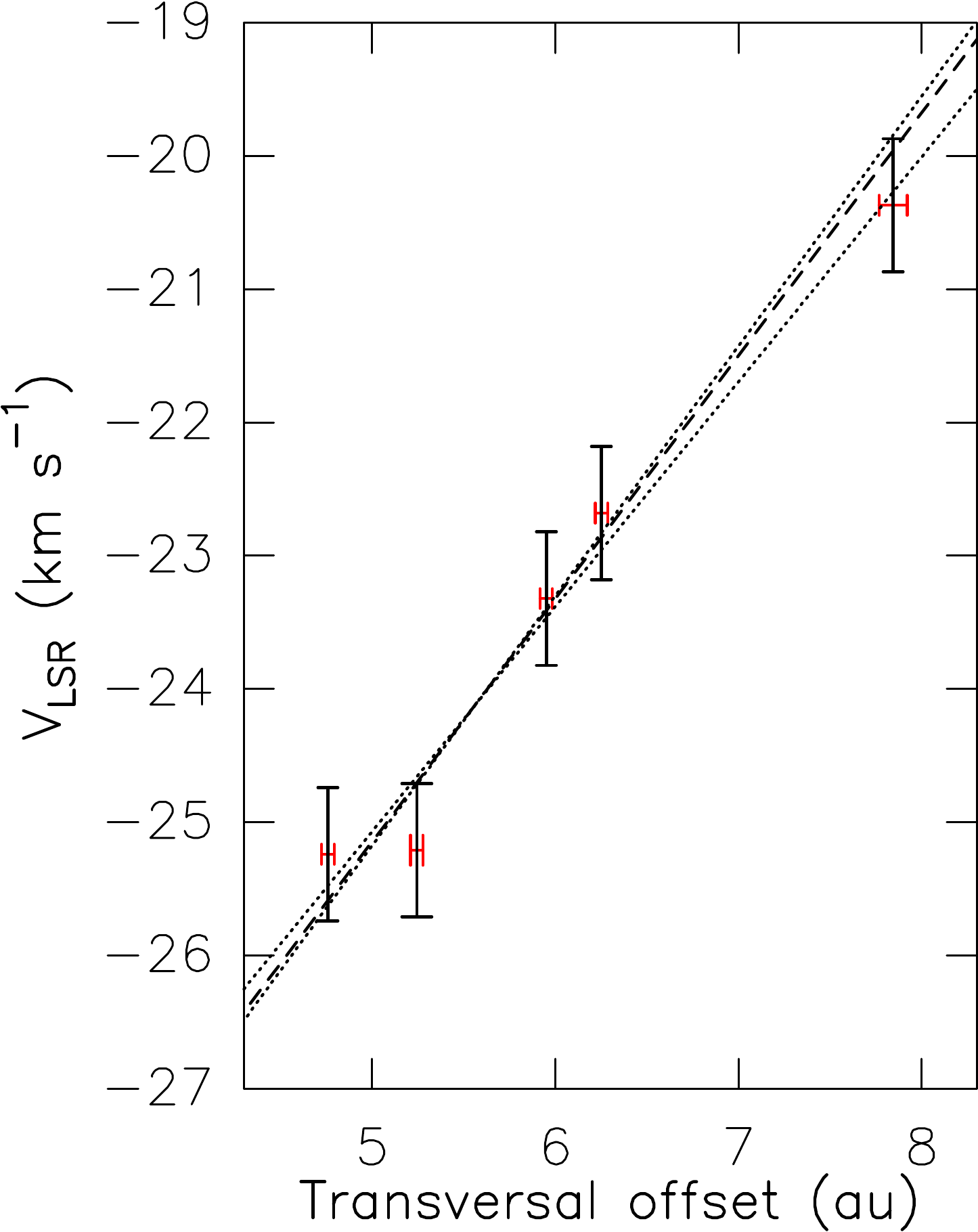}
\caption{Estimate of \ $\omega$ \ at the base of the NE-1 stream. Plot of \Vlsr \ versus \ $R$ \ for the maser cluster at the base of the NE-1 stream, inside the dashed rectangle in Fig.~\ref{NE_spi}a. Vertical  black and horizontal red errorbars indicate velocities and radii (with corresponding errors), respectively. The black dashed and dotted lines show the best linear fit and the associated uncertainty, respectively.}
\label{NE-1-root_1}
\end{figure*}
\begin{figure*}%
\centering
\includegraphics[width=0.65\textwidth]{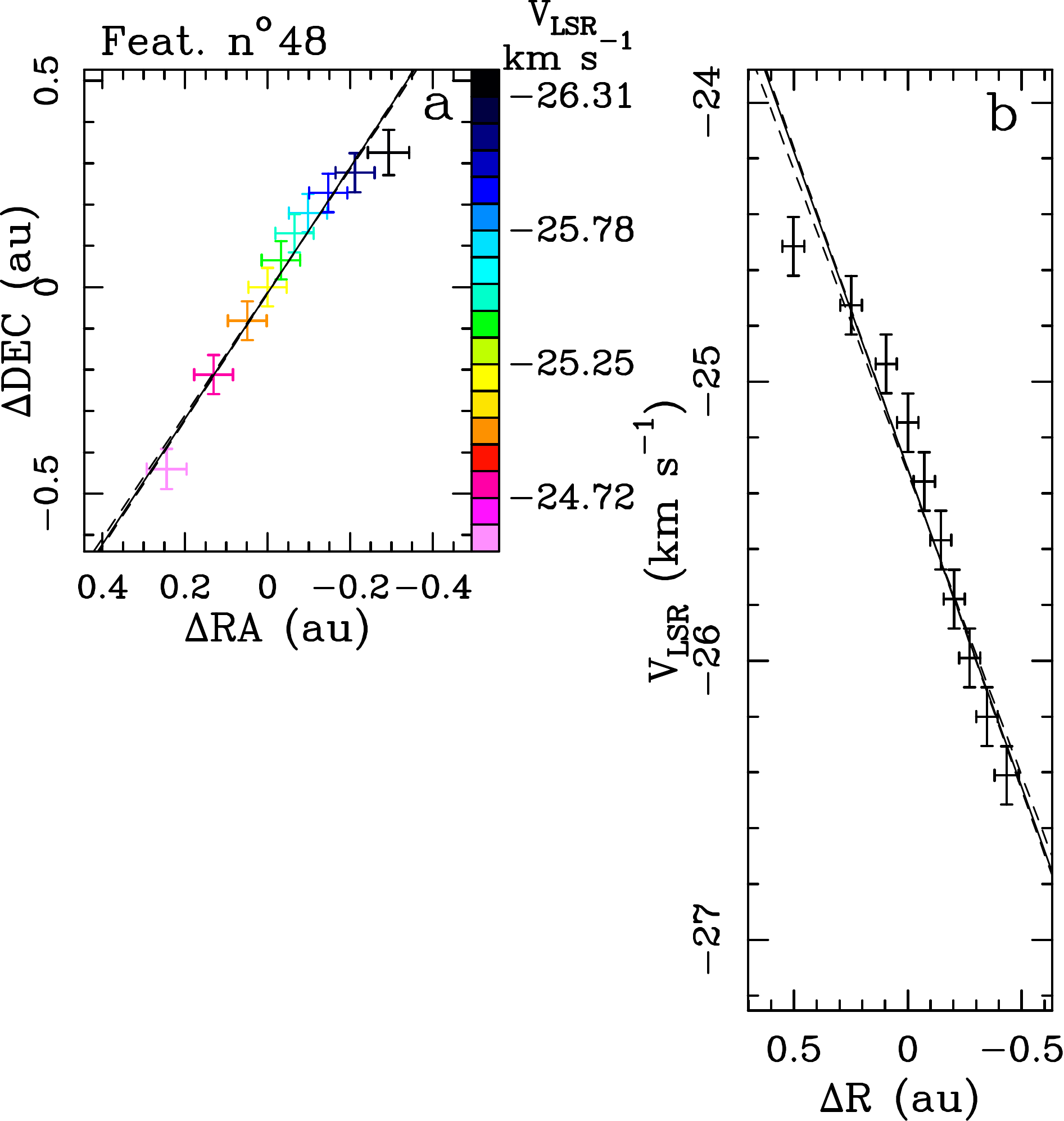}
\caption{Rotation inside individual maser features. (a)~Colored errorbars give positions (and corresponding errors) of the individual spots collected into the maser feature, with colors denoting the maser \Vlsr. The black continuous and dashed lines give the best linear fit of the spot positions and the associated uncertainty, respectively. \ (b)~Plot of the spot \Vlsr\ versus positions projected along the major axis of the spot distribution. The black continuous and dashed lines show the best linear fit of the plotted values and the associated uncertainty, respectively.}
\label{NE-1-root_2}
\end{figure*}

Fig.~\ref{NE-1-root_1} shows that the \Vlsr\ of the five masers at the base ($z \approx$~20~au) of the NE-1 stream (inside the dashed rectangle of Fig.~\ref{NE_spi}a) changes linearly with \ $R$. The derived value of \ $\omega = 1.82\pm0.09$~\kmsau \ is significantly higher than the value, $1.15\pm0.09$~\kmsau (see Table~\ref{tab_fit}), determined at larger elevations. That can be interpreted as evidence that the gas closer to the Alfv\'en point rotates faster than that at higher elevations, in agreement with the MHD disk-wind theory. As described in Appendix~\ref{met_obs}, each maser feature is a collection of many compact spots for which we have accurately measured position and \Vlsr. For the maser feature (out of the five) nearest to the jet axis, Fig.~\ref{NE-1-root_2}a shows that the spatial distribution of the spots is linear and elongated close to (7\degree away from) the disk axis, and Fig.~\ref{NE-1-root_2}b shows that the spot \Vlsr\ changes linearly with the position, $\Delta R$, along the major axis of the spatial distribution. Given its orientation, it is natural to interpret the \Vlsr\ gradient internal to the maser feature in terms of rotation, and from the linear fit of \Vlsr\ versus $\Delta R$ \ we derive a value of \ $\omega = 2.3\pm0.1$~\kmsau. This should still be considered a lower limit for \ $\omega_{\rm K}$, the angular velocity at the Alfv\'en point.  Beside the one just considered, other two maser features present a well defined, internal \Vlsr\ gradient. In these two features, the linear distribution of the spots is directed at larger angles, 20--30\degree, from the disk axis, and the amplitudes of the \Vlsr\ gradients are  \ $1.8\pm0.1$ \ and \ $3.1\pm0.1$~\kmsau. These two features have larger separation from the jet axis and the corresponding larger inclination of the spot distribution with the plane of rotation could indicate that the internal \Vlsr\ gradient is influenced by non-rotation terms, as, for instance, the increase of the streaming velocity with the elevation. In any case, assuming that also for these two features the internal  \Vlsr\ gradient is dominated by rotation, we obtain a consistent average value for the in-feature angular velocity of  \ $\omega =$~2.4~\kmsau.
 
\begin{figure*}
\centering
\includegraphics[width=0.7\textwidth]{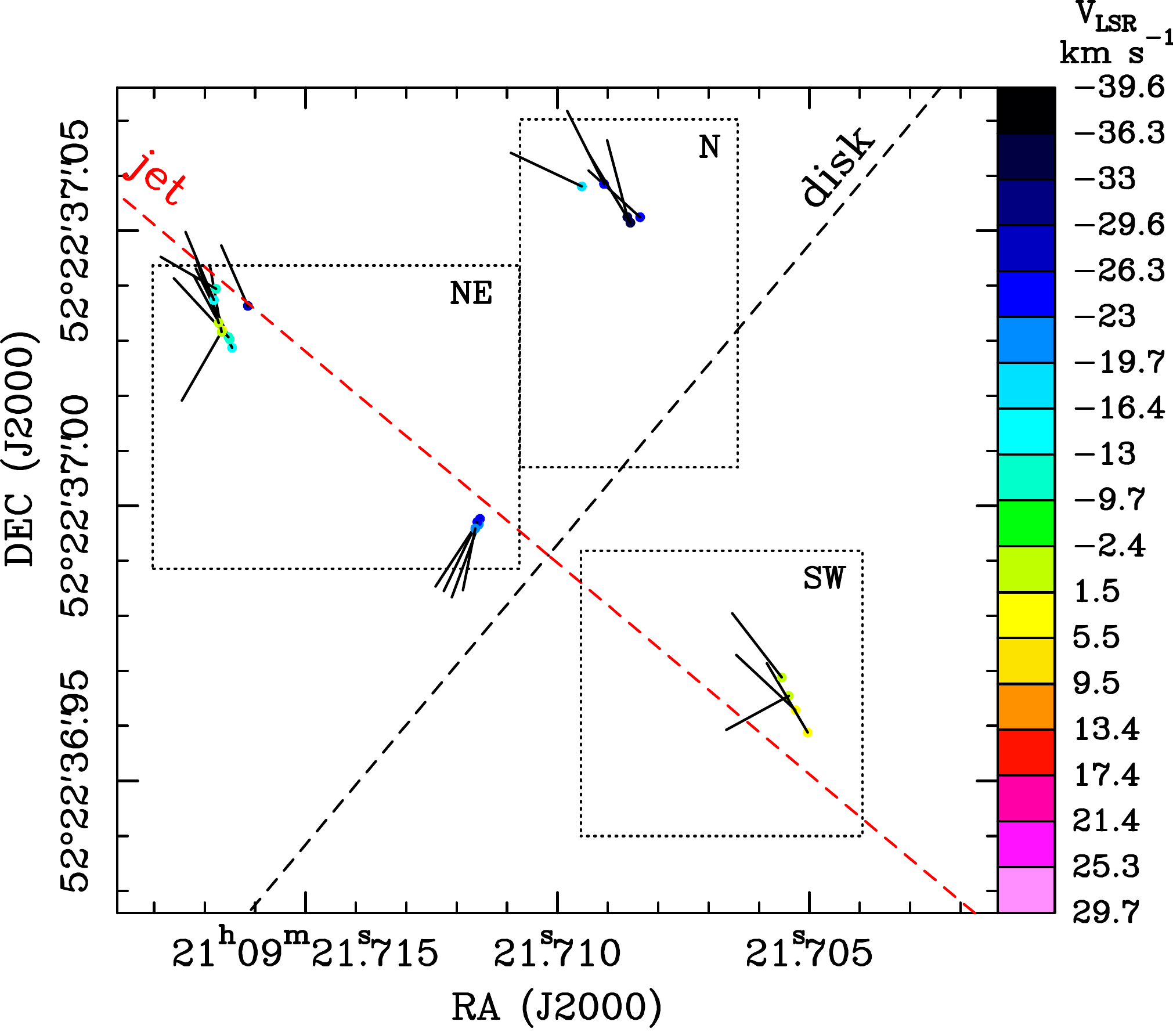}
\caption{Spatial distribution of internal velocity gradients of 22~GHz water maser features. Colored dots, black dotted rectangles, and the red and black dashed lines have the same meaning as in Fig.~\ref{glo}a. The black segments show the velocity gradient orientation. Only the maser features which yield correlation coefficients  \ $\ge$~0.7 \ from the linear fits of both the spot positions (see Fig.~\ref{NE-1-root_2}a) and \Vlsr\ versus projected positions (see Fig.~\ref{NE-1-root_2}b), are plotted.}
\label{grad}
\end{figure*}

We stress that the internal motion of the maser features does not contribute to the features' \Vlsr\ used in our analysis of the maser velocity pattern on larger scales (see Figs.~\ref{SW_spi},~\ref{NE_spi}~and~\ref{N_wire}). In fact, the feature \Vlsr \ is obtained by averaging the internal spots' \Vlsr \ (see Appendix~\ref{met_obs}), with the result that any contribution from internal motions is averaged out.
In the previous analysis, we have interpreted the maser internal motions, on scales of a few au, as a product of the larger scale (10--100~au) motions. Since we have considered only five features within a small area, this interpretation has to be better justified. Fig.~\ref{grad} shows the internal gradient orientation for all the maser features in \targ\ with well defined internal gradients. Most of the gradients are directed close to either the jet or disk axis. That suggests that the maser internal motions trace mainly the large scale dynamics, dominated by either the streaming flow motion or rotation. Interestingly, a previous study\cite{Mos11b} of a sample of 6.7~GHz methanol maser sources reached a similar conclusion. 
Finally, we note that, on the size (a few au) of the maser features, turbulent motions have to be significantly smaller than the typical maser line widths or  $\le$~0.1~\kms, otherwise they would destroy the velocity coherence along the maser amplification path.

\section{Kinematics of the water masers further to NE and S}
\label{met_NES}

In the main text, we have considered only the filaments of maser emission inside the NE, SW and N regions, but more scattered maser features are also observed further to NE close to the jet axis and S of the YSO (see Fig.~\ref{glo}a).
The outmost northeastern features are found approximately in the same region of the previous VLBA maser detections (see Fig.~\ref{glo}a). However the newly observed masers are fewer than before and their \Vlsr\ is now within \ $\pm$5~\kms \ from $V_{\rm sys}$ instead of covering the range \ [$-$38, 0]~\kms \ as before. The directions of the VLBA maser proper motions are well collimated but their amplitudes (in the range \ [10, 50]~\kms) scatter significantly in position (see Fig.~\ref{NV}b), which, as discussed in Appendix~\ref{met_shock}, suggests that these masers arise in external shocks and do not trace the true flow speed. In our previous analysis\cite{Mos21}, we have interpreted these outmost northeastern masers as external shocks at the wall of a cylindrical MHD jet interacting with the surrounding medium. Applying the equations for a MHD disk wind, we have derived a terminal jet speed of \ $\approx$~200~\kms \ and a launch radius of \ $\approx$~2~au. These values agree with the trend of the calculations reported in Table~\ref{tab_der}, where higher streaming velocities correspond to smaller launch radii.

The southern features are distributed in three separated clusters (see Fig.~\ref{glo}a): \ C1)~$z \le 20$~au \ and \ 9.5~\kms \ $\le$ \Vlsr\ $\le$ 17.4~\kms; 
\ C2)~$20  \le z \le 100$~au \ and \ 5.5~\kms \ $\le$ \Vlsr\ $\le$ 9.5~\kms;
\ C3)~$z \ge 120$~au \ and \ 5.5~\kms \ $\le$ \Vlsr\ $\le$ 29.7~\kms. Comparing the C1~\&~C2 clusters with the N region, maser positions and \Vlsr\ are about symmetrical with respect to the YSO; beside, each of the two clusters presents an arc-like shape similar to the N streamline. Thus, it is possible that these clusters trace small portions of two co-rotating streamlines emerging at different disk radii. Finally, the maser cluster C3 has position and \Vlsr\ approximately symmetrical to the emission previously observed with the VLBA further to NE (see Fig.~\ref{NV}b), which suggests that masers in C3 can emerge in shocks at the wall of the southeastern lobe of the jet.

\section{Simulation snapshot of a forming massive star}
\label{met_simu}

\begin{figure*}
	\includegraphics[width=\textwidth]{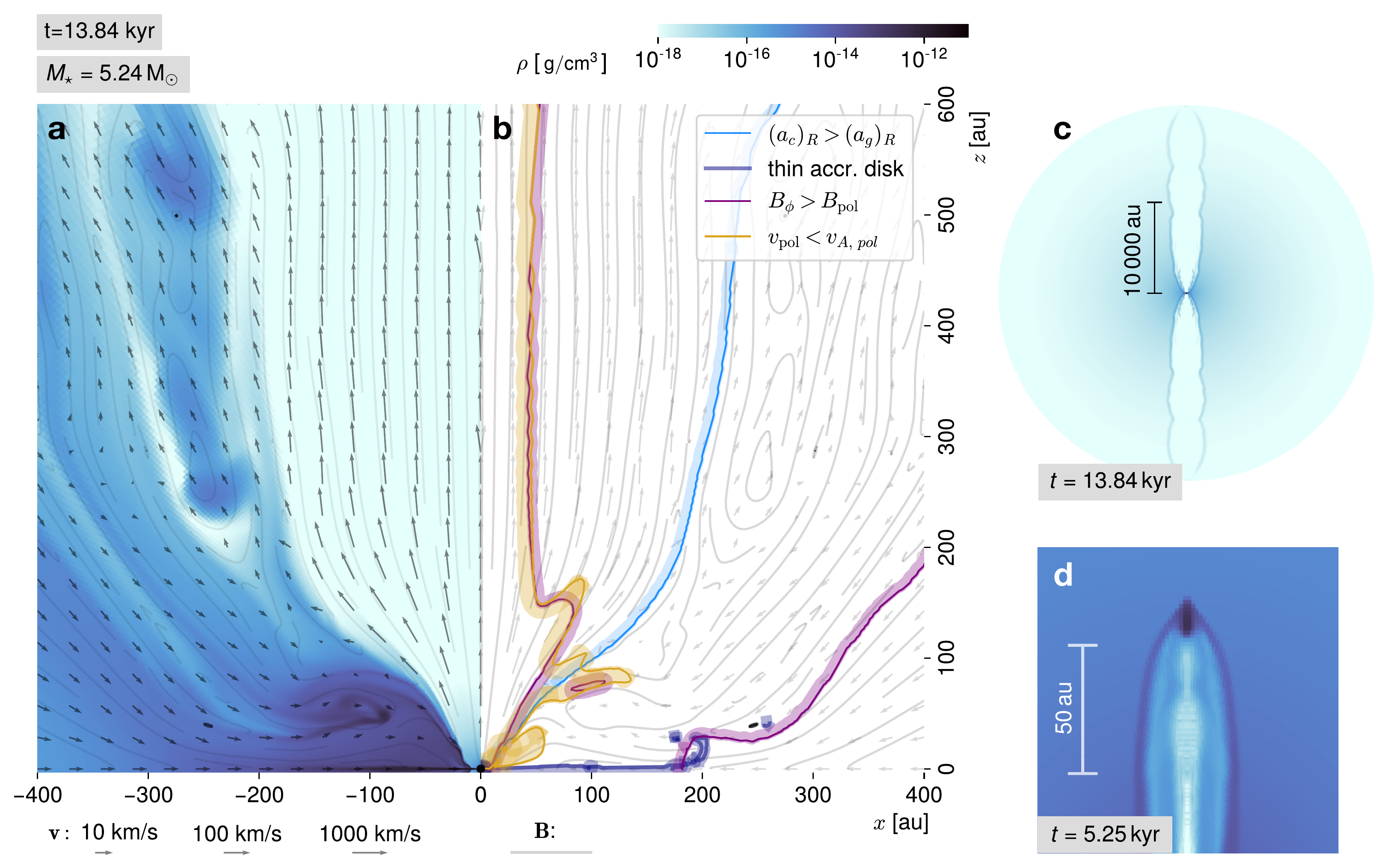}
	\caption{Simulation snapshot of a forming massive star, surrounded by an accretion disk and 
	magnetically-driven outflows. \ (a)~Density, velocity and magnetic field lines. \ (b)~Contours that show 
	the regions where: the material is sub-Alfv\'enic (yellow), the magnetic field lines are wound by rotation (purple), 
	and the centrifugal acceleration is stronger than gravity (blue). 
	The shadow that accompanies each contour indicates the region where the inequality listed in the legend holds true. 
	The contour lines have been smoothened in order to filter out local and short-lived features. \ (c)~Large-scale 
	structure of the outflows. \ (d)~Bow-shock produced in the simulation.}
	\label{fig: simulation-results}
\end{figure*}

As part of a more extensive study we will describe in a forthcoming article, we performed an axisymmetrical
 simulation of the formation of a massive star starting from the gravitational collapse of a rotating cloud
 core threaded by a magnetic field (GAO, RK et al., in preparation). We used the methods of magnetohydrodynamics to model the weakly ionized
 gas and dust with the code Pluto\cite{Mig07}, with an Ohmic resistivity model as a non-ideal 
 effect\cite{Mac07}, and additional modules for self-gravity\cite{Kui10} and the transport of the
 thermal radiation emitted by the gas and dust\cite{Kui20}.

The cloud core has an initial mass of \ 100~\ms\ and a radius of \ 0.1~pc. 
The assumed conditions for the onset of the gravitational collapse ($t=0$) are as follows: the density is 
distributed according to \ $\rho \propto r^{-1.5}$, the cloud core rotates like a solid body with a rotational 
energy equivalent to 4\% of its gravitational energy content, and the magnetic field is uniform. The initial 
magnitude of the magnetic field is determined by the mass-to-flux ratio, which we take as 20 times the critical 
(collapse-preventing) value\cite{Mou76} and corresponds to a relatively weak initial 
magnetic field. A constant value of the opacity of \ 1~cm$^{2}$~g$^{-1}$ \ was used to model the gas and 
dust, as well as an initial dust-to-gas mass ratio of 1\%.

We used an axisymmetrical grid of 896$\times$160 cells in spherical coordinates, with the radial coordinate
increasing logarithmically with the distance to the center of the cloud. An inner boundary of
 \ 3~au \ was set up, inside of which the protostar is formed through accretion. 
 No flows are artificially injected from the inner boundary into the collapsing cloud.

The simulation starts with an initial gravitational collapse epoch. After $\sim$~5~kyr, enough 
angular momentum is transported to the center of the cloud to start forming an accretion disk that grows 
in size over time. Roughly at the same time, we observe the launch of magnetically-driven outflows. 
Magnetic pressure arising from the dragging of magnetic field lines by the rotating flow eventually
overcomes gravity and seeds the formation of the outflow cavity, thrusting a bow shock in the process 
(Fig.~\ref{fig: simulation-results}d), which propagates outwards as the cavity grows in size. 
Previous observations of \targ\ have uncovered the presence of a bow shock located at distances 
of \ $\approx$~36000~au from the forming massive YSO\cite{Mos21}. The initial launch of the magnetically-driven outflows 
provides a possible formation mechanism for the observed bow shock and in return, the propagation of the 
bow shock provides an estimation for the age of the system.

In the simulation, the protostar reaches a mass of \ 5.24~\ms \ 
(a value in the expected mass interval from observations) after \ 13.84~kyr \ of evolution. 
We estimate that the bow shock has propagated to a distance of $\approx$~30000~au \ at that time, 
roughly in line with the observations. At the same time, the accretion disk has grown to 
about \ 180~au \ in radius, in agreement with the observational estimates\cite{Mos21}, as well. 
The data reveal a magneto-centrifugally launched jet, in a similar way as reported by the literature \cite{Bla82,Koe18}, 
however, we see that the launching region of the jet is narrowed by the ram pressure of the infalling material 
from the envelope and the presence of a thick layer of the accretion disk which is vertically supported 
by magnetic pressure. Material transported from large scales through the accretion disk reaches 
the launching region, located at \ $z\lesssim 100$~au (see Fig.~\ref{fig: simulation-results}b), 
where the centrifugal force is stronger than gravity, and simultaneously, 
where the flow becomes sub-Alfv\'enic in the co-rotating frame and the magnetic field lines are mostly poloidal. 
Under those conditions, a parcel of plasma is accelerated along the field lines until the flow becomes 
super-Alfv\'enic again, a point at which the magnetic field lines become mostly toroidal. 
This causes the parcel to acquire a helical trajectory (depicted previously 
in Fig.~\ref{glo}b). The geometry and kinematics of the region where the flow becomes 
helical in the simulation coincides roughly with the positions and velocities of the 
observed masers of the NE and SW regions. 
At larger rotation radii, the broader magnetic field lines present in the outflow cavity could 
give rise to the kinematic footprint detected in the masers in the N region.

At distances of \ $\sim$~10000~au, where the outflow material propagates through 
the cloud (Fig.~\ref{fig: simulation-results}c), we notice the existence of re-collimation zones 
that arise because of magnetic hoop stress and ram pressure from the envelope. 
The position of the re-collimation zones is similar to the zones where lobes of possible synchrotron 
emission have been previously observed around the YSO \cite{Mos21}, thus providing a possible mechanism for 
their formation.

We provide this comparison as an example that a magneto-centrifugally launched jet around a 
forming massive star yields a consistent picture with the observations. However, the coincidences should be 
taken with caution, as different combinations of initial conditions may yield similar results, 
which means that a wider and deeper investigation is needed in order to determine the conditions of the 
onset of star formation from the observational data.

\end{appendix}

\clearpage

\onecolumn 

{ \large \bf \noindent Correspondence and Request for Material:}

\vspace{0.3cm} Any request should be addressed to Luca Moscadelli

\vspace{1cm} {\large \bf \noindent Data Availability:}

\vspace{0.3cm} This letter makes use of the following EVN data: GM077 (EVN project code). 

The datasets generated during and/or analyzed during the current study are available from the corresponding author on reasonable request.

\vspace{1cm} {\large \bf \noindent Author Contribution Statement:}

\vspace{0.3cm} L.M. led the project, analysis, discussion, and drafted the manuscript.

A.S, H.B., and R.K. commented on the manuscript and participated in the discussion.
 
G.A.O. and R.K. performed the numerical jet simulations described in Appendix~\ref{met_simu}.

G.A.O. performed the dynamical analysis of the simulations, compared the simulation results to the observations, and produced the illustrations of the magnetic field lines and streamlines.

\vspace{1cm} {\large \bf \noindent Competing Interests:
}

\vspace{0.3cm} The authors declare no competing (financial and non-financial) interests.


\end{document}